\documentstyle{ar}
\input epsf.tex
\def\abs#1{\left| #1\right|}
\def\mdot{\hskip -.1cm \cdot \hskip -.1cm}
\newcommand{\dsp}{\displaystyle}
\def\abs#1{\left| #1\right|}

\let\badcite=\cite
\def\cite{~\badcite}
\catcode`@=11
\def\ifundefined#1{\expandafter\ifx\csname#1\endcsname\relax}
\def\citenum#1{\ifundefined{b@#1}{\bf#1}%
   \immediate\write16{citenum: Undefined argument #1}%
   \else\csname b@#1\endcsname\fi}
\catcode`@=12

\def\slashchar#1{\setbox0=\hbox{$#1$}           
   \dimen0=\wd0                                 
   \setbox1=\hbox{/} \dimen1=\wd1               
   \ifdim\dimen0>\dimen1                        
      \rlap{\hbox to \dimen0{\hfil/\hfil}}      
      #1                                        
   \else                                        
      \rlap{\hbox to \dimen1{\hfil$#1$\hfil}}   
      /                                         
   \fi}                                         %

\def\dofig#1#2{%
   \vskip-12pt%
   \centerline{\epsfxsize=#1\epsfbox{#2}}%
   \vskip-12pt}

\def\anp#1,#2(#3){{\it Adv.\ Nucl.\ Phys.\ }{\rm #1}: {\rm#2} {\rm(#3)}}
\def\aip#1,#2(#3){{\it Am.\ Inst.\ Phys.\ }{\rm #1}: {\rm#2} {\rm(#3)}}
\def\aj#1,#2(#3){{\it Astrophys.\ J.\ }{ #1}: {\rm#2} {\rm(#3)}}
\def\ajs#1,#2(#3){{\it Astrophys.\ J.\ Supp.\ }{\rm #1}: {\rm#2} {\rm(#3)}}
\def\ajl#1,#2(#3){{\it Astrophys.\ J.\ Lett.\ }{\rm #1}: {\rm#2} {\rm(#3)}}
\def\ajp#1,#2(#3){{\it Am.\ J.\ Phys.\ }{\rm #1}: {\rm#2} {\rm(#3)}}
\def\apny#1,#2(#3){{\it Ann.\ Phys.\ (NY)\ }{\rm #1}: {\rm#2} {\rm(#3)}}
\def\apnyB#1,#2(#3){{\it Ann.\ Phys.\ (NY)\ }{\rm B#1}: {\rm#2} {\rm(#3)}}
\def\apD#1,#2(#3){{\it Ann.\ Phys.\ }{\rm D#1}: {\rm#2} {\rm(#3)}}
\def\ap#1,#2(#3){{\it Ann.\ Phys.\ }{\rm #1}: {\rm#2} {\rm(#3)}}
\def\ass#1,#2(#3){{\it Ap.\ Space Sci.\ }{\rm #1}: {\rm#2} {\rm(#3)}}
\def\astropp#1,#2(#3)%
    {{\it Astropart.\ Phys.\ }{\rm #1}: {\rm#2} {\rm(#3)}}
\def\aap#1,#2(#3)%
    {{\it Astron.\ \& Astrophys.\ }{\rm #1}: {\rm#2} {\rm(#3)}}
\def\araa#1,#2(#3)%
    {{\it Ann.\ Rev.\ Astron.\ Astrophys.\ }{\rm #1}: {\rm#2} {\rm(#3)}}
\def\arnps#1,#2(#3)%
    {{\it Ann.\ Rev.\ Nucl.\ and Part.\ Sci.\ }{\rm #1}: {\rm#2} {\rm(#3)}}
\def\arns#1,#2(#3)%
   {{\it Ann.\ Rev.\ Nucl.\ Sci.\ }{\rm #1}: {\rm#2} {\rm(#3)}}
\def\cqg#1,#2(#3){{\it Class.\ Quantum Grav.\ }{\rm #1}: {\rm#2} {\rm(#3)}}
\def\cpc#1,#2(#3){{\it Comp.\ Phys.\ Comm.\ }{\rm #1}: {\rm#2} {\rm(#3)}}
\def\cjp#1,#2(#3){{\it Can.\ J.\ Phys.\ }{\rm #1}: {\rm#2} {\rm(#3)}}
\def\cmp#1,#2(#3){{\it Commun.\ Math.\ Phys.\ }{\rm #1}: {\rm#2} {\rm(#3)}}
\def\cnpp#1,#2(#3)%
   {{\it Comm.\ Nucl.\ Part.\ Phys.\ }{\rm #1}: {\rm#2} {\rm(#3)}}
\def\cnppA#1,#2(#3)%
   {{\it Comm.\ Nucl.\ Part.\ Phys.\ }{\rm A#1}: {\rm#2} {\rm(#3)}}
\def\el#1,#2(#3){{\it Europhys.\ Lett.\ }{\rm #1}: {\rm#2} {\rm(#3)}}
\def\epjC#1,#2(#3){{\it Europhys.\ J.\ }{\rm C#1}: {\rm#2} {\rm(#3)}}
\def\grg#1,#2(#3){{\it Gen.\ Rel.\ Grav.\ }{\rm #1}: {\rm#2} {\rm(#3)}}
\def\hpa#1,#2(#3){{\it Helv.\ Phys.\ Acta }{\rm #1}: {\rm#2} {\rm(#3)}}
\def\ieeetNS#1,#2(#3)%
    {{\it IEEE Trans.\ }{\rm NS#1}: {\rm#2} {\rm(#3)}}
\def\IEEE #1,#2(#3)%
    {{\it IEEE }{\rm #1}: {\rm#2} {\rm(#3)}}
\def\ijar#1,#2(#3)%
  {{\it Int.\ J.\ of Applied Rad.\ } {\rm #1}: {\rm#2} {\rm(#3)}}
\def\ijari#1,#2(#3)%
  {{\it Int.\ J.\ of Applied Rad.\ and Isotopes\ } {\rm #1}: {\rm#2} {\rm(#3)}}
\def\jcp#1,#2(#3){{\it J.\ Chem.\ Phys.\ }{\rm #1}: {\rm#2} {\rm(#3)}}
\def\jgr#1,#2(#3){{\it J.\ Geophys.\ Res.\ }{\rm #1}: {\rm#2} {\rm(#3)}}
\def\jetp#1,#2(#3){{\it Sov.\ Phys.\ JETP\ }{\rm #1}: {\rm#2} {\rm(#3)}}
\def\jetpl#1,#2(#3)%
   {{\it Sov.\ Phys.\ JETP Lett.\ }{\rm #1}: {\rm#2} {\rm(#3)}}
\def\jpA#1,#2(#3){{\it J.\ Phys.\ }{\rm A#1}: {\rm#2} {\rm(#3)}}
\def\jpG#1,#2(#3){{\it J.\ Phys.\ }{\rm G#1}: {\rm#2} {\rm(#3)}}
\def\jpamg#1,#2(#3)%
    {{\it J.\ Phys.\ A: Math.\ and Gen.\ }{\rm #1}: {\rm#2} {\rm(#3)}}
\def\jpcrd#1,#2(#3)%
    {{\it J.\ Phys.\ Chem.\ Ref.\ Data\ } {\rm #1}: {\rm#2} {\rm(#3)}}
\def\jpsj#1,#2(#3){{\it J.\ Phys.\ Soc.\ Jpn.\ }{\rm G#1}: {\rm#2} {\rm(#3)}}
\def\lnc#1,#2(#3){{\it Lett.\ Nuovo Cimento\ } {\rm #1}: {\rm#2} {\rm(#3)}}
\def\nature#1,#2(#3){{\it Nature} {\rm #1}: {\rm#2} {\rm(#3)}}
\def\nc#1,#2(#3){{\it Nuovo Cimento} {\rm #1}: {\rm#2} {\rm(#3)}}
\def\nim#1,#2(#3)%
   {{\it Nucl.\ Instrum.\ Methods\ }{\rm #1}: {\rm#2} {\rm(#3)}}
\def\nimA#1,#2(#3)%
    {{\it Nucl.\ Instrum.\ Methods\ }{\rm A#1}: {\rm#2} {\rm(#3)}}
\def\nimB#1,#2(#3)%
    {{\it Nucl.\ Instrum.\ Methods\ }{\rm B#1}: {\rm#2} {\rm(#3)}}
\def\np#1,#2(#3){{\it Nucl.\ Phys.\ }{\rm #1}: {\rm#2} {\rm(#3)}}
\def\mnras#1,#2(#3){{\it MNRAS\ }{\rm #1}: {\rm#2} {\rm(#3)}}
\def\medp#1,#2(#3){{\it Med.\ Phys.\ }{\rm #1}: {\rm#2} {\rm(#3)}}
\def\mplA#1,#2(#3){{\it Mod.\ Phys.\ Lett.\ }{\rm A#1}: {\rm#2} {\rm(#3)}}
\def\npA#1,#2(#3){{\it Nucl.\ Phys.\ }{\rm A#1}: {\rm#2} {\rm(#3)}}
\def\npB#1,#2(#3){{\it Nucl.\ Phys.\ }{\rm B#1}: {\rm#2} {\rm(#3)}}
\def\npBps#1,#2(#3){{\it Nucl.\ Phys.\ (Proc.\ Supp.) }{\rm B#1}:
{\rm#2} {\rm(#3)}}
\def\pasp#1,#2(#3){{\it Pub.\ Astron.\ Soc.\ Pac.\ }{\rm #1}: {\rm#2} {\rm(#3)}}
\def\pl#1,#2(#3){{\it Phys.\ Lett.\ }{\rm #1}: {\rm#2} {\rm(#3)}}
\def\fp#1,#2(#3){{\it Fortsch.\ Phys.\ }{\rm #1}: {\rm#2} {\rm(#3)}}
\def\ijmpA#1,#2(#3)%
   {{\it Int.\ J.\ Mod.\ Phys.\ }{\rm A#1}: {\rm#2} {\rm(#3)}}
\def\ijmpE#1,#2(#3)%
   {{\it Int.\ J.\ Mod.\ Phys.\ }{\rm E#1}: {\rm#2} {\rm(#3)}}
\def\plA#1,#2(#3){{\it Phys.\ Lett.\ }{\rm A#1}: {\rm#2} {\rm(#3)}}
\def\plB#1,#2(#3){{\it Phys.\ Lett.\ }{\rm B#1}: {\rm#2} {\rm(#3)}}
\def\pnasus#1,#2(#3)%
   {{\it Proc.\ Natl.\ Acad.\ Sci.\ \rm (US)}{B#1}: {\rm#2} {\rm(#3)}}
\def\ppsA#1,#2(#3){{\it Proc.\ Phys.\ Soc.\ }{\rm A#1}: {\rm#2} {\rm(#3)}}
\def\ppsB#1,#2(#3){{\it Proc.\ Phys.\ Soc.\ }{\rm B#1}: {\rm#2} {\rm(#3)}}
\def\pr#1,#2(#3){{\it Phys.\ Rev.\ }{\rm #1}: {\rm#2} {\rm(#3)}}
\def\prA#1,#2(#3){{\it Phys.\ Rev.\ }{\rm A#1}: {\rm#2} {\rm(#3)}}
\def\prB#1,#2(#3){{\it Phys.\ Rev.\ }{\rm B#1}: {\rm#2} {\rm(#3)}}
\def\prC#1,#2(#3){{\it Phys.\ Rev.\ }{\rm C#1}: {\rm#2} {\rm(#3)}}
\def\prD#1,#2(#3){{\it Phys.\ Rev.\ }{\rm D#1}: {\rm#2} {\rm(#3)}}
\def\prept#1,#2(#3){{\it Phys.\ Reports\ } {\rm #1}: {\rm#2} {\rm(#3)}}
\def\prslA#1,#2(#3)%
   {{\it Proc.\ Royal Soc.\ London }{\rm A#1}: {\rm#2} {\rm(#3)}}
\def\prl#1,#2(#3){{\it Phys.\ Rev.\ Lett.\ }{\rm #1}: {\rm#2} {\rm(#3)}}
\def\ps#1,#2(#3){{\it Phys.\ Scripta\ }{\rm #1}: {\rm#2} {\rm(#3)}}
\def\ptp#1,#2(#3){{\it Prog.\ Theor.\ Phys.\ }{\rm #1}: {\rm#2} {\rm(#3)}}
\def\ppnp#1,#2(#3)%
        {{\it Prog.\ in Part.\ Nucl.\ Phys.\ }{\rm #1}: {\rm#2} {\rm(#3)}}
\def\ptps#1,#2(#3)%
   {{\it Prog.\ Theor.\ Phys.\ Supp.\ }{\rm #1}: {\rm#2} {\rm(#3)}}
\def\pw#1,#2(#3){{\it Part.\ World\ }{\rm #1}: {\rm#2} {\rm(#3)}}
\def\pzetf#1,#2(#3)%
   {{\it Pisma Zh.\ Eksp.\ Teor.\ Fiz.\ }{\rm #1}: {\rm#2} {\rm(#3)}}
\def\rgss#1,#2(#3){{\it Revs.\ Geophysics \& Space Sci.\ }{\rm #1}:
        {\rm#2} {\rm(#3)}}
\def\rmp#1,#2(#3){{\it Rev.\ Mod.\ Phys.\ }{\rm #1}: {\rm#2} {\rm(#3)}}
\def\rnc#1,#2(#3){{\it Riv.\ Nuovo Cimento\ } {\rm #1}: {\rm#2} {\rm(#3)}}
\def\rpp#1,#2(#3)%
    {{\it Rept.\ on Prog.\ in Phys.\ }{\rm #1}: {\rm#2} {\rm(#3)}}
\def\science#1,#2(#3){{\it Science\ } {\rm #1}: {\rm#2} {\rm(#3)}}
\def\sjnp#1,#2(#3)%
   {{\it Sov.\ J.\ Nucl.\ Phys.\ }{\rm #1}: {\rm#2} {\rm(#3)}}
\def\panp#1,#2(#3)%
   {{\it Phys.\ Atom.\ Nucl.\ }{\rm #1}: {\rm#2} {\rm(#3)}}
\def\spu#1,#2(#3){{\it Sov.\ Phys.\ Usp.\ }{\rm #1}: {\rm#2} {\rm(#3)}}
\def\surveyHEP#1,#2(#3)%
    {{\it Surv.\ High Energy Physics\ } {\rm #1}: {\rm#2} {\rm(#3)}}
\def\yf#1,#2(#3){{\it Yad.\ Fiz.\ }{\rm #1}: {\rm#2} {\rm(#3)}}
\def\zetf#1,#2(#3)%
   {{\it Zh.\ Eksp.\ Teor.\ Fiz.\ }{\rm #1}: {\rm#2} {\rm(#3)}}
\def\zp#1,#2(#3){{\it Z.~Phys.\ }{\rm #1}: {\rm#2} {\rm(#3)}}
\def\zpA#1,#2(#3){{\it Z.~Phys.\ }{\rm A#1}: {\rm#2} {\rm(#3)}}
\def\zpC#1,#2(#3){{\it Z.~Phys.\ }{\rm C#1}: {\rm#2} {\rm(#3)}}
\def\etal{\hbox{\it et~al.}}

\begin{document}
\baselineskip=14pt

\title{ The QCD Coupling Constant \footnotemark} 
\footnotetext{This work was supported in part by the Director, Office of
Science, Office of High Energy and Nuclear Physics, Division of High Energy
Physics of the U.S. Department of Energy under Contracts
DE-AC03-76SF00098.}
\author{Ian Hinchliffe\affiliation{Lawrence Berkeley National
    Laboratory, Berkeley, CA} and Aneesh V.~Manohar\affiliation{Physics 
    Department, University of California at San
  Diego, 9500 Gilman Drive, La Jolla, CA 92093-0319}}
\begin{keywords}
QCD,
\end{keywords} 
\begin{abstract}
       This paper presents a summary of the current status of
       determinations of the strong coupling constant $\alpha_s$.
A detailed description of the definition, scale dependence and
inherent theoretical ambiguities is given.
The various physical processes that can be used to determine
$\alpha_s$ are reviewed and attention is given to the uncertainties,
both theoretical and experimental.
\end{abstract}
\maketitle
%
\newpage 

\def\Dsl{\hbox{/\kern-.6000em D}} 
\def\dsl{\,\raise.15ex\hbox{/}\mkern-13.5mu D} 
\def\tr{{\rm tr\,}}
\def\msbar{$\overline{\rm MS}$}
\newcommand{\bra}[1]{{\left\langle{#1}\right|}}
\newcommand{\ket}[1]{{\left|{#1}\right\rangle}}
\def\vev#1{\left\langle{#1}\right\rangle}
\newcommand{\me}[3]{{\left\langle{#1}\vphantom{#2 #3}
\right|{#2}\left|\vphantom{#1 #2}{#3}\right\rangle}}
\newcommand{\rme}[3]{{\left\langle{#1}\vphantom{#2 #3}
\right.\parallel{#2}\left.\parallel\vphantom{#1 #2}{#3}\right\rangle}}
\def\im{{\rm Im \,}}

\section{QCD AND ITS COUPLING}

Quantum chromodynamics (QCD) is a gauge field theory that describes the strong
interactions of quarks and gluons \cite{politzer}. All experimental results to date are
consistent with QCD predictions to within the experimental and theoretical
errors. In this review, we discuss the current status of the extraction of the
strong interaction coupling constant $\alpha_s$ from the experimental data.

The QCD Lagrangian describing the interactions of quarks and gluons is
\begin{equation}\label{1}
L = -{1 \over 4} F_{\mu \nu}^a F^{a\mu \nu} + \sum_k \overline \psi_k \left( 
i \dsl - m_k \right) \psi_k,
\end{equation}
where
\begin{equation}
F_{\mu \nu}^a = \partial_\mu A_\nu^a -  \partial_\nu A_\mu^a+ g f^{abc}
A_\mu^b A_\nu^c
\end{equation}
is the gluon field strength tensor,
\begin{equation}
D_\mu = \partial_\mu - i g A_\mu^a T^a
\end{equation}
is the gauge covariant derivative, and $T^a$ are the $SU(3)$ representation
matrices normalized so that $\tr T^a T^b = \delta^{ab}/2$, and the sum on $k$ is
over the six different flavors ($u,d,s,c,b,t$) of quarks. At the classical
level, the QCD Lagrangian depends on the six quark masses $m_k$, and the strong
interaction coupling constant $g$, or equivalently, the strong fine-structure
constant $\alpha_s=g^2/4\pi$. The quantum theory contains an additional
parameter, the $\theta$-angle, that violates CP. The experimental limit on this
parameter is $\theta < 10^{-9}$ \cite{theta}, so we will set it to zero for the purposes of
this article.

One can evaluate QCD scattering amplitudes in powers of $\alpha_s$ using a
Feynman diagram expansion. As is typical in a quantum field theory, loop graphs
are divergent and need to be treated using a renormalization scheme. The most
commonly used scheme is modified minimal subtraction  (\msbar) \cite{msbar}, and we will
use this scheme throughout. An important consequence of renormalization is that
the parameters $\alpha_s$ and $m_k$ of the QCD Lagrangian depend in a
calculable manner on the \msbar\ subtraction-scale $\mu$. The $\mu$ dependence
of $\alpha_s$ is described by the $\beta$-function,
\begin{equation}\label{3}
\mu{d \alpha_s \over d \mu} = \beta\left (\alpha_s \left(\mu \right)\right).
\end{equation}
In perturbation theory,
\begin{equation}
\beta\left (\alpha_s \right) = -{\beta_0} {\alpha_s^2 \over 2 \pi } - 
{\beta_1}{\alpha_s^3 \over \left(2 \pi \right)^2} -
{\beta_2}{\alpha_s^4 \over \left(2 \pi \right)^3} - \ldots,
\end{equation}
where (for $n_f$ flavors of quarks)
\begin{eqnarray}
\beta_0=11-{2 \over 3} n_f,\\
\beta_1=51-{19 \over 3} n_f,
\end{eqnarray}
and the next two terms are also known~\cite{larin97}.

If $\alpha_s$ is small, the renormalization group equation
Eq.~(\ref{3}) can be integrated using only the $\beta_0$ term to give
\begin{equation}\label{af}
{1\over \alpha_s(\mu_1)} = {1\over \alpha_s(\mu_1)} + {\beta_0 \over 2 \pi}
 \ln{\mu_1 \over \mu_2}.
\end{equation}
Since $\beta_0>0$ for $n_f < 16.5$, $\alpha_s(\mu) \to 0$ as $\mu \to \infty$.
The vanishing of the QCD coupling for large values of $\mu$ is referred to as
asymptotic freedom.  One important consequence of asymptotic freedom, is that
QCD processes at high energies can be reliably computed in a perturbation
expansion in $\alpha_s$.

A measurable quantity,
such as the total cross-section for $e^+ e^- \to {\rm hadrons}$ at high energies
can be computed as a function of the QCD coupling constant $\alpha_s(\mu)$ and
the center of mass energy $E_{\rm CM}$,
\begin{equation}\label{2}
\sigma\left(e^+ e^- \to {\rm hadrons}\right)={1\over E_{\rm CM}^2}
f\left(\alpha_s\left(\mu\right),\ln{E_{\rm CM} \over \mu} \right).
\end{equation}
where $f$ is a dimensionless function of its arguments. The form of the
cross-section given in Eq.~(\ref{2}) follows from dimensional analysis: $\sigma$
has dimensions of ${\rm energy}^{-2}$, and $\mu$ has dimensions of energy. In
the \msbar\ sscheme, any dependence on $\mu$ is logarithmic, so $f$ can only
depend on $\ln E_{\rm CM}/\mu$. The cross section 
$\sigma\left(e^+ e^- \to {\rm hadrons}\right)$ is a measurable quantity and
cannot depend on the subtraction scale $\mu$, so the $\mu$ dependence on the
right hand side of Eq.~(\ref{2}) must cancel,
\begin{equation}\label{5}
\mu {d \over d \mu} 
f\left(\alpha_s\left(\mu\right),{E_{\rm CM} \over \mu} \right)=0,
\label{eq:mu}
\end{equation}
and any value of $\mu$ can be used on the right hand side of Eq.~(\ref{2}).
In practice, one can only compute the right hand side of Eq.~(\ref{2}) at some
finite order in perturbation theory, and the approximate value of
$f$ can depend on $\mu$ at higher order in perturbation theory. Typically, one
finds that the perturbation expansion has terms of the form
\begin{equation}
\left[ \alpha_s(\mu) \ln {E_{\rm CM} \over \mu} \right]^n,
\end{equation}
which are referred to as ``leading logarithms.'' Even if $\alpha_s(\mu)$ is
small, the perturbation expansion can break down if $\ln {E_{\rm CM}/\mu}$ is
large. For this reason, it is conventional to choose the subtraction scale of
order the center of mass energy $E_{\rm CM}$. The exact choice of scale (for
example, whether $\mu=2E_{\rm CM}$ or $E_{\rm CM}$ or $E_{\rm CM}/2$) is
arbitrary, and differences in choice of scale are formally of higher order in
$\alpha_s$. Many methods have been proposed to determine the optimum scale to
use for a given calculation \cite{choice}. The only way to determine the ``best'' scale at a
given order is to compute the cross-section at next order. [Of course, in this
case, one might as well use the more accurate formula to determine the
cross-section.] The scale dependence of a given quantity can also be used to
estimate the size of neglected higher order corrections. Scale dependence is a
dominant source of error in many of the quantities that will be used to
determine $\alpha_s$.

Perturbation theory is valid if one chooses $\mu$ to be of order $E_{\rm CM}$,
so that the expansion parameter is $\alpha_s(E_{\rm CM})$, with no large
logarithms. This shows that at high-energies, the coupling constant is small
because of asymptotic freedom, and QCD cross-sections are approximately those
of free quarks and gluons. At low, energies, non-perturbative effects become
important.

The value of $\alpha_s$ is determined by computing a quantity in terms of
$\alpha_s$, and comparing with its measured value. One might think that it is
better to use high-energy processes to determine $\alpha_s$, since perturbation
theory is more reliable. This is not necessarily the case. High energy
processes can be computed more reliably precisely because they do not depend
very much on $\alpha_s$. This means that errors in the experimental measurement
or theoretical calculation get amplified when they are converted to an error on
$\alpha_s$. We will see in this article that low-energy extractions of
$\alpha_s$ have comparable errors to  those at high energy.

In addition to scale dependence, the coupling constant $\alpha_s$ is also
subtraction scheme dependent. The scheme dependence of the coupling constant is
compensated for by the scheme dependence in the functional form for a
measurable quantity, so that the value of an observable is scheme independent.
The \msbar\ scheme will be used in this article, but there is still some
residual scheme dependence we need to consider. In the \msbar\ scheme, heavy
quarks do not decouple in loop graphs at low-energy. For example, the
$\beta$-function coefficient $\beta_0=11-2n_f/3$, where $n_f$ is the number of
quark flavors. This expression is true for all energies, irrespective of the
mass of the quark. One might expect that at energies much smaller than the mass
$m_Q$ of a heavy quark, the quark does not contribute to the $\beta$-function.
This  cannot happen in the \msbar\ scheme, since \msbar\ is a mass-independent
subtraction scheme, which results in a mass-independent $\beta$-function. What
happens is that there are large logarithms of the form $\ln m_Q/\mu$ that
compensate for the ``incorrect'' $\beta$-function at low-energies. In practice,
one deals with this problem using an effective field theory. At energies
smaller than $m_Q$, one switches from a QCD Lagrangian with $n_f$ flavors to a
QCD Lagrangian with $n_f-1$ flavors by integrating out the heavy quark flavor.
The effect of the heavy quark is taken into account by higher dimension
operators in the QCD Lagrangian, and by shifts in the parameters $\alpha_s$ and
$m_k$. Thus it is necessary to specify the effective theory when quoting the
value of $\alpha_s$. We will use the notation $\alpha_s^{(n_f)}$ to denote the
value of $\alpha_s$ in the $n_f$ flavor theory. One can compute the relation
between $\alpha_s^{(n_f)}$ and $\alpha_s^{(n_f-1)}$ at the scale $\mu=m_Q$ of
the heavy quark. This relation is known to three loops~\cite{chetyk97},
\begin{eqnarray}
\alpha_s^{(n_f-1)}\left(m_Q\right) &=& \alpha_s^{(n_f)}\left(m_Q\right)\left[
1 + 0.1528 \left( {\alpha_s^{(n_f)}\left(m_Q\right) \over \pi }\right)^2
\nonumber \right.\\
&&\left.+ (0.9721- 0.0847 n_f ) \left( {\alpha_s^{(n_f)}\left(m_Q\right) \over \pi }
\right)^3 + \ldots \right].\label{7}
\end{eqnarray}
When quoting the value of $\alpha_s(\mu)$, it is also necessary to specify
the number of flavors in the effective theory. In most of the $\alpha_s$
determinations we consider, the appropriate effective theory to use is one with
$n_f=5$, and $\alpha_s$ will refer to $\alpha_s^{(5)}$ unless otherwise
specified.

Classical QCD is a scale invariant theory, but this scale invariance is broken
at the quantum level. The quantum theory has a dimensionful parameter $\Lambda$
that characterizes the scale of the strong interactions. The $\Lambda$ parameter
is determined in terms of $\alpha_s(\mu)$. The solution of the renormalization
group equation Eq.~(\ref{3}) including the first three terms in the
$\beta$-function is
\begin{eqnarray}\label{6}
\alpha_s(\mu)&=&{4 \pi \over \beta_0 \ln \left( \mu^2 / \Lambda^2 \right)} \left[
1 - {2 \beta_1 \over \beta_0^2} {\ln \left[ \ln \left( \mu^2 / \Lambda^2 \right)
\right] \over \ln \left( \mu^2 / \Lambda^2 \right) } \right. \nonumber \\
&&\left. +{4 \beta_1^2 \over \beta_0^4 \ln^2 \left( \mu^2 / \Lambda^2 \right)}
\left( \left( \ln \left[ \ln \left( \mu^2 / \Lambda^2 \right) \right]-{1\over 2}
\right)^2 + {\beta_2 \beta_0 \over \beta_1^2} - {5 \over 4} \right) \right].
\end{eqnarray}
This equation can be used to determine $\Lambda$ if $\alpha_s$ is known at some
scale $\mu$. The value of $\Lambda$ depends on the number of terms retained in
Eq.~(\ref{6}). The expansion parameter in Eq.~(\ref{6}) is
\begin{equation}
{\ln \left[ \ln \left( \mu^2 / \Lambda^2 \right)
\right] \over \ln \left( \mu^2 / \Lambda^2 \right) },
\end{equation}
which is small as long as $\mu \gg \Lambda$. In QCD, $\Lambda$ is of order
200
~MeV. The last term in Eq.~(\ref{6}) is often dropped in the definition of
$\Lambda$. For a fixed value of $\alpha_s(M_Z)$, the shift in $\Lambda$ is
approximately 15~MeV if the last term in Eq.~(\ref{6}) is dropped. 

The QCD $\beta$-function depends on $n_f$, and so changes across quark
thresholds. This in turn implies that $\Lambda$ changes across quark thresholds,
so that $\Lambda^{(n_f)}$ is the value of $\Lambda$ with $n_f$ dynamical quark
flavors. The matching conditions for $\Lambda^{(n_f)} \to \Lambda^{(n_f-1)}$
can be computed using the matching condition Eq.~(\ref{7}) for $\alpha_s$. The
differences between $\Lambda^{(3)}$, $\Lambda^{(4)}$, and $\Lambda^{(5)}$ are
numerically very significant.

In addition to the perturbative effects discussed so far, non-perturbative
effects play an important role in strong interaction processes. The size of
non-perturbative effects is governed by the ratio of the strong interaction
scale $\Lambda$ to the typical energy $E_{\rm CM}$ of a given process. In many
cases, non-perturbative effects are estimated using a model analysis, or by a
phenomenological fit to the experimental data. A few processes can be analyzed
rigorously using the operator product expansion (OPE)\cite{ope}; in this case, one can
calculate the size of non-perturbative effects in terms of matrix elements of
gauge invariant local operators. Two classic examples of this type are deep
inelastic scattering, and the total cross-section for $e^+ e^- \to {\rm
hadrons}$. 

It is convenient to calculate $R(s)$, the ratio of the total
cross-sections for $e^+ e^- \to {\rm hadrons}$ and $e^+ e^- \to \mu^+ \mu^-$ at
center of mass-energy $E_{\rm CM}=\sqrt{s}$.
One can show that
\begin{equation}
R(s)= f_0 \left( \alpha_s(s)\right) + f_1 \left( \alpha_s(s)\right) {
\vev{{F_{\mu \nu }F ^{\mu \nu}}} \over s^2}
+ \ldots,
\end{equation}
where the first non-perturbative correction depends on the vacuum expectation
value of the square of the gluon field-strength tensor. By dimensional
analysis, this quantity is of order $\Lambda^4$, so the non-perturbative
corrections are of order $\Lambda^4/s^2$. The existence of an OPE provides some
crucial information on the size of non-perturbative corrections. For $R(s)$, we
know that the corrections vanish at least as fast as $\Lambda^4/s^2$ for large
values of $s$, because $F_{\mu \nu }F ^{\mu \nu}$ is the lowest dimension
operator that can contribute in the OPE. Similarly, it is known in deep
inelastic scattering that the first non-perturbative corrections arise from
twist-four operators, and are of order $\Lambda^2/Q^2$, where $Q$ is the
momentum transfer. One can estimate the size of non-perturbative corrections
for processes with an OPE, by estimating the value of operator matrix elements.
In some cases, one is fortunate enough that the relevant matrix element can
actually be determined from some other measurement, or computed from first
principles. The size of non-perturbative corrections is much less certain if
the process does not have an OPE. Non-perturbative effects can fall off like a
fractional power of $\Lambda/s$, or could have some more complicated dependence
on $s$. Typically, one uses some model estimate of the non-perturbative
corrections.

\begin{figure}
\epsfxsize=5cm
\hfil\epsfbox{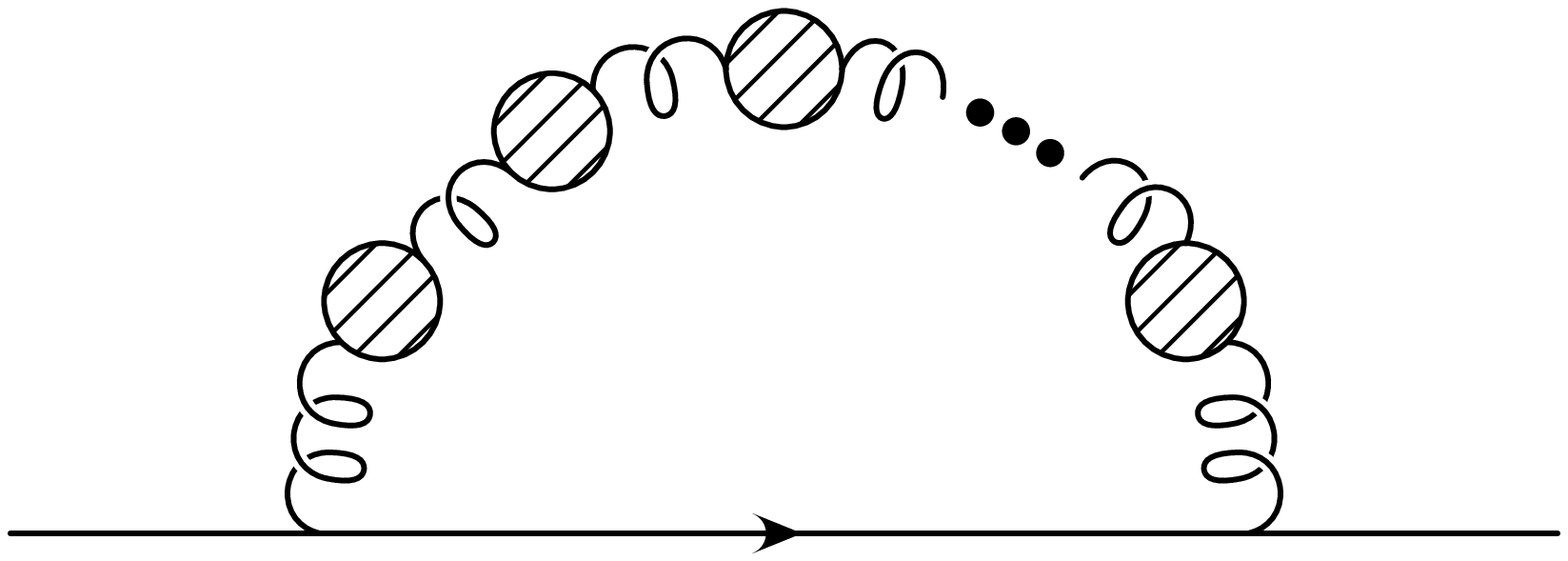}\hfill
\caption{Graphs that contribute to the propagator correction
in the bubble chain approximation.
\label{fig:bubble}}
\end{figure}

Perturbative and non-perturbative corrections to scattering cross-sections are
interrelated, because the QCD perturbation series is an asymptotic expansion,
rather than a convergent expansion. A dimensionless quantity $f(\alpha_s)$ has
an expansion of the form
\begin{equation}
f(\alpha_s) = c_0 + c_1 \alpha_s + c_2 \alpha_s^2 + \ldots.
\end{equation}
Typically, the coefficients $c_n$ grow as $n!$, so that the series has zero
radius of convergence. The large-order behavior of the perturbation series can
be computed in certain limiting
cases~\cite{lautrup,thooft,david,bsuv,beneke,ren-review}. If one studies QCD in the limit of a
large number of flavors,
$n_f \to \infty$, with $\alpha_s n_f$ fixed, one can sum all terms of the form
$(\alpha_s n_f)^n$. This is sometimes referred to as the ``bubble chain''
approximation, because the graphs one sums are of the type show in
Fig.~\ref{fig:bubble}.
QCD is not asymptotically free as $n_f \to \infty$. Nevertheless, one can try
and apply the bubble chain results to QCD. The bubble chain graphs contribute
to the QCD $\beta$-function. In the large $n_f$ limit, the coefficient $\beta_0
= 11 -2 /3 n_f \to -2/3 n_f$. One therefore computes the bubble chain sum,
makes the replacement $n_f \to -3 \beta_0/2$, and uses the resultant expression
for QCD with
$\beta_0 > 0$. This seemingly unjustified procedure has provided some useful
insights into the nature of the QCD perturbation series. A detailed discussion
of this method is beyond the scope of the present article. It is typically found
that the coefficients $c_n$ in the perturbation expansion have a factorial
divergence in the bubble chain approximation. One can try and sum a series of
this type using a Borel transformation. One defines the Borel transform of $f$
by
\begin{equation}
f_B(t) = c_0 \delta (t) + c_1 + c_2 t + \ldots + {c_{n+1}\over n!} t^n + \ldots.
\end{equation}
Then the original function can be obtained by the inverse Borel transform,
\begin{equation}
f(\alpha_s) = \int_0^\infty dt\ f_B(t) e^{-t/\alpha_s}.
\end{equation}
Suppose the coefficients of $f(\alpha_s)$ have the form
\begin{equation}
c_{n+1} = a^n n! ,\ \ n > 0.
\end{equation}
Then
\begin{equation}
f_B(t) = c_0 \delta (t) + \sum_{n=0}^\infty  a^n t^n = c_0 \delta(t) + {1\over 1
- a t},
\end{equation}
and the inverse Borel transform gives
\begin{equation}
f(\alpha_s) = c_0  +  \int_0^\infty dt\  e^{-t/\alpha_s} {1\over 1
- a t}.
\end{equation}
The behavior of the integral is governed by the singularities in the complex $t$
plane, which are referred to as renormalons. If $a<0$, the integral is
well-defined, since the singularity at $t=1/a$ is not along the path of
integration. If $a>0$, the singularity is along the contour of integration. One
can regulate the integral by deforming the contour around the singularity. The
integral depends on the precise prescription used. The prescription dependence
is related to the pole at $t=1/a$,
which gives a contribution to the integral of the form
\begin{equation}
\exp\left[-{1 \over a\alpha_s}\right]= \exp\left[{-\beta_0 \ln(\mu/\Lambda) \over
2 \pi a} \right] = \left( {\Lambda \over \mu} \right)^{\beta_0 / 2 \pi a}.
\end{equation}
The value of $\mu$ is chosen to be the typical energy in the process, such as
the momentum transfer $Q$ for deep inelastic scattering. The perturbative series
then has the same structure as a typical non-perturbative correction, a power
law correction of the form
\begin{equation}\label{8}
\left( {\Lambda^2 \over Q^2} \right)^{u_0}
\end{equation}
where $u_0=\beta_0/(4 \pi a)$ is the renormalon singularity in the variable
$u=\beta_0 t/(4 \pi)$.  [It is conventional to refer to the location of
the renormalon singularity in $u$ rather than in $t$.] Contributions of the form
Eq.~(\ref{8}) are called renormalon ambiguities, since their value depends on
the way in which one performs the inverse Borel transform.

Renormalon ambiguities have the same structure as non-perturbative corrections.
It has been suggested in the literature \cite{ren-review} that renormalons can be used as a guide
to the size of non-perturbative corrections. There is one non-trivial check to
this idea. A renormalon singularity at $u_0$ corresponds to a non-perturbative
ambiguity of the form Eq.~(\ref{8}). In processes that have an OPE, all
non-perturbative effects should be given by the matrix elements of gauge
invariant local operators. To every renormalon ambiguity in the perturbation 
expansion, there should be a corresponding ambiguity in the operator matrix
element, such that the sum is well-defined~\cite{david}. This can only happen provided that
there is a gauge invariant local operator corresponding to every renormalon
singularity. For example, the first gauge invariant operator corrections to
deep inelastic scattering are of the form $\Lambda^2/Q^2$, and it is known that
the first renormalon ambiguity is at $u=1$. Similarly, for $R$, the first
non-perturbative corrections are of the form $\Lambda^4/Q^4$, and the first
renormalon ambiguity is at $u=2$. The matching between renormalon singularities
and the OPE occurs in all examples that have been computed so far. For this
reason, renormalon singularities have also been taken as an indication of the
size of non-perturbative effects in processes without an OPE. Non-perturbative
effects are expected to fall off faster if the renormalons are at larger values
of $u$.

Infrared sensitivity is also used to estimate the size of non-perturbative
corrections to a measurable quantity~\cite{akhoury}. One computes the quantity
in the presence of an infrared cutoff momentum $\lambda$. For example, one can
imagine working in a box of size $1/\lambda$, or using a gluon mass of order
$\lambda$. Cross-sections can have infrared divergences of the form $\ln
\lambda$. If one  computes measurable cross-sections for color singlet states
to scatter into color singlet states, one finds that the $\ln \lambda$ terms
cancel, and the cross-section is infrared finite, a result known as the KLN
\cite{KLN} theorem. It is important to include finite detector resolution to get a finite
cross-section, as for QED. For example, the Bhabha scattering cross-section for
$e^+ e^- \to e^+ e^-$  has an infrared divergence at one-loop order. However,
it is impossible to distinguish  $e^+ e^- \to e^+ e^-$ from $e^+ e^- \to e^+
e^- \gamma $ if the photon energy $E_\gamma$ is smaller than the detector
resolution $\delta$. The measurable quantity is the sum of the $e^+ e^- \to e^+
e^-$  cross-section and the $e^+ e^- \to e^+ e^- \gamma $ cross-section for
$E_\gamma < \delta$, which is free of infrared singularities. While the $\ln
\lambda$ term must cancel, terms of order $\lambda$, $\lambda^2 \ln \lambda$,
etc.\ which vanish as $\lambda \to 0$ need not cancel. The first non-vanishing
term is an indication of the infrared-sensitivity of a given quantity. In QCD,
one can imagine that the scale $\lambda$ represents the confinement scale
$\Lambda$. A process that is infrared sensitive at order $\lambda^n$ would then
be expected to have non-perturbative corrections of order $\Lambda^n/Q^n$,
where $Q$ is the typical momentum transfer. In a few cases, one can analyze the
problem using renormalon methods, and by using the criterion of infrared
sensitivity. It is found in these cases that both methods give the same
estimate for the size of non-perturbative corrections.

\section{$\alpha_s$ FROM $Z$ DECAYS AND $e^+e^-$ TOTAL RATES}

The total cross section for $e^+ e^- \to \hbox{hadrons}$
is obtained (at low values of $\sqrt s$)
by multiplying the muon-pair cross section by the
factor $R$. At lowest order in QCD perturbation theory $R=R^0 =
3\Sigma_q e_q^2 $ where $e_q$ is the electric charge of the quark of
flavor $q$.  The
higher-order QCD corrections
to this are known, and the results can be
expressed in terms of the factor:
\begin{equation}
        R = R^{(0)}
                \left [ 1 +
                           \frac{\alpha_s}{\pi}
                         +  C_2 
                  \left (
                        \frac{\alpha_s}{\pi}
                  \right )^2
                         +  C_3
                  \left (
                        \frac{\alpha_s}{\pi}
                  \right )^3
                         +   \;\cdots\;\;
                \right ]
\label{Rfactor}
\end{equation}
where $C_2=1.411$ and $C_3=-12.8$\cite{Gorishny91}.

This result is only correct in the zero-quark-mass limit.
The ${\cal O}(\alpha_s $) corrections
are also known for massive quarks\cite{kuhn94}.
The principal advantages 
of determining $\alpha_s$ from~$R$ in $e^+ e^-$ annihilation
are  that the measurement is inclusive, that there is no dependence
on the details of the hadronic final state and that non-perturbative
corrections are suppressed by $1/s^2$.

A measurement by CLEO \cite{cleoR} at $\sqrt{s}=10.52$ GeV yields
$
\alpha_s ( 10.52  \hbox{ GeV} ) = 0.20 \pm 0.01 \pm 0.06
$  which corresponds to $
\alpha_s ( M_Z ) = 0.13 \pm 0.005 \pm 0.03.
$
A comparison of the theoretical
prediction of Eqn.~\ref{Rfactor}\ (corrected for the $b$-quark mass),
with all the available data at values of $\sqrt s$  between 20 and 65~GeV,
gives\cite{haidt}
$
\alpha_s ( 35 \hbox{ GeV} ) = 0.146 \pm 0.030.
$ 
It should be noted that the size of the order $\alpha_s^3$ term is of order 
40\% of that of the order
$\alpha_s^2$ and 3\% of the order $\alpha_s$. 
If the order $\alpha_s^3$ term is
not included, the extracted value decreases to $\alpha_s ~( 35 \hbox{ GeV} )
 = 0.142 \pm 0.03$, a difference smaller than the experimental
 error.

Measurements of the ratio of the hadronic to leptonic width of the $Z$ at LEP
and SLC,  
$\Gamma_h/\Gamma_{\mu}$ probe the same quantity as
$R$. Using the average of $\Gamma_h/\Gamma_{\mu}=20.783 \pm 0.029$
gives
$\alpha_s(M_Z)=0.123\pm 0.004$\cite{lepz}.
The prediction depends upon the couplings of the quarks and leptons to 
the $Z$. The precision is such that higher order electroweak
corrections to these couplings must be included.
There are theoretical errors arising from 
the values of top-quark and Higgs masses which enter in these
radiative corrections. 
Hence, while this method has small theoretical uncertainties
from QCD itself, it relies sensitively on the electroweak
couplings of the $Z$ to quarks\cite{blondel93} and on the ability of
the Standard Model of electroweak interactions to predict these correctly.  
The presence of new physics which changes these couplings via electroweak
radiative corrections would invalidate the extracted value of
$\alpha_s(M_Z)$.
Since the Standard Model fits the measured $Z$ properties well, this
concern is ameliorated and
more precise value of $\alpha_s$ can be obtained
by using a global fit to the many
precisely measured properties of the $Z$ boson and the  measured 
$W$ and top masses.
This gives \cite{langsports} 
$$\alpha_s(M_Z)=0.1192\pm 0.0028$$
This error is larger than the shift in the value of $\alpha_s(M_Z)$
($\sim 0.002$) that would result if the order $\alpha_s(M_Z)^3$ term
were omitted and hence one can conclude that it is very unlikely that the
uncertainty due to the 
unknown $\alpha_s(M_Z)^4$ terms will dominate over the experimental
uncertainty.

\section{DETERMINATION OF $\alpha_s$ FROM DEEP INELASTIC SCATTERING}

\label{sec:struc}

\begin{figure}
\epsfxsize=5cm
\hfil\epsfbox{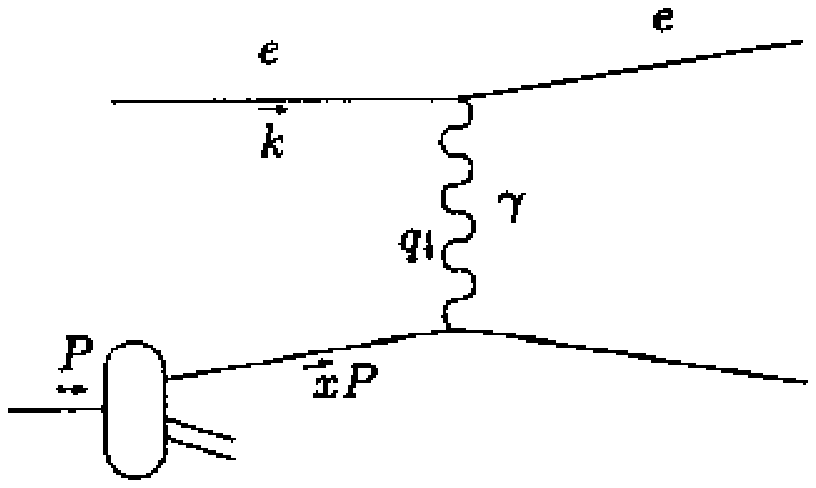}\hfill
\caption{Figure showing the kinematics of deep inelastic scattering
  $ep\to eX$
\label{fig:refd}}
\end{figure}

The original and still one 
of the most powerful quantitative tests
of perturbative QCD is the breaking of Bjorken scaling in deep-inelastic
lepton-hadron scattering.  
 Consider the case of electron-proton
scattering ($ep\to eX$), where the cross-section can be written as
\begin{equation}
\frac{d\sigma}{dxdy} =\frac{4\pi\alpha_{em}^2s}{Q^4}
\left [\frac{1+(1-y)^2}{2}2xF_1(x,Q^2) + (1-y)(F_2(x,Q^2)-
2xF_1(x,Q^2))\right]
\label{eq19}
\end{equation}
The variables are defined as follows (see Figure \ref{fig:refd}):  $q$ is the momentum
of the exchanged photon, $P$ is the momentum of the target proton, $k$
is that of the incoming electron, and
\begin{eqnarray}
Q^2&=-q^2\nonumber \\
\nu&= \frac{q\cdot P}{m_p}
\nonumber \\
x&=\frac{Q^2}{2m_p\nu}\nonumber \\
y&=\frac{q\cdot p}{k\cdot p}\nonumber \\
s&=2p\cdot k+m_p^2\nonumber \\
\end{eqnarray}

For charged current scattering, which proceeds via the exchange of a virtual
$W$ boson between the lepton and target nucleus, there is an additional parity
violating structure function $F_3$
\begin{equation}
\begin{array}{ll}
\dsp \frac{d\sigma^{\nu N}}{dxdy}=&\dsp\frac{G_F^2M_W^4s}
{2\pi (Q^2+M_W^2)^2}(xy^2F_1^{\nu N}(x,Q^2)\cr
&\dsp +(1-y-x^2  y^2M^2/Q^2)F_2^{\nu N}(x,Q^2)\cr
& \dsp -\frac{1}{2}x((1-y)^2-1)F_3^{\nu N}(x,Q^2))
\end{array}
\end{equation}
For $\overline{\nu}N$ scattering 
the sign of the last ($xF_3$) term is reversed.

In the leading-logarithm
approximation, the measured structure functions $F_i (x,  Q^2$) are
related to the quark distribution functions $q_i (x, Q^2 )$
according to the naive
parton model, for example
\begin{equation}
F_2(x,Q^2)=\sum_i e_i^2 q_i (x, Q^2 )
\label{eq:parton}
\end{equation}
  Here $q_i (x, Q^2 )$ is the probability for a parton of type $i$ to
  carry a fraction $x$ of the nucleon's momentum.
The $Q^2$ dependence of the parton distribution functions is predicted 
by perturbative QCD, hence a measurement of the $Q^2$ dependence
(``scaling violation'') can by used to measure $\alpha_s$.
In describing the way in which scaling is broken in QCD, it is
convenient to define nonsinglet and singlet quark distributions:
$$
         F^{NS}  =  q_i  -  q_{j} \qquad \qquad
         F^S  =  \sum_i ( q_i  +  \overline{q}_i )
\label{FNS}
$$
The nonsinglet structure functions have nonzero values of flavor quantum
numbers such as isospin or baryon number.
The variation with $Q^2$ of these is
described by the so-called
DGLAP equations\cite{Altarelli77}\cite{altpar77}:
$$
\begin{array}{ll}
            Q^2\;
                \frac{\partial F^{NS}}{\partial Q^2} & =
                \frac{\alpha_s ( | Q |  )}{2 \pi}
                     P^{qq} \ast F^{NS} \\
\label{AltarelliParisi;a}
                                        \cr
             Q^2\;
                 \frac{\partial}{\partial Q^2}
                       \left(  {F^S\atop G}  \right)   & =
                  \frac{\alpha_s ( | Q |  )}{2 \pi}
            \left(
                       {P^{qq}\phantom{{}_q}\atop P^{gq\phantom{{}^Y}}}\;
                        {2 n_f P^{qg}\atop P^{{gg}\phantom{{}^Y}}}
            \right)
                         \ast \left(   {F^S\atop G}  \right) 
\label{AltarelliParisi;b}
\end{array}$$
where $\ast$ denotes a convolution integral:
$$
           f \ast  g = \int_x^1
                 \frac{dy}{y}\;
               f (y)\; g \left (\frac{x}{y}\right ) 
\label{fstarg}
$$
The leading-order Altarelli-Parisi\cite{altpar77} splitting functions are
$$
\begin{array}{ll}
               P^{qq} & = 
                        \frac{4}{3}
                \left [
                        \frac{1 +  x^2}{(1 - x)_+}
                \right ]
                         +  2 \delta ( 1 - x )
\label{APsf;a}
\cr
               P^{qg} & =
                        \frac{1}{2}
                        \left [ x^2  +  ( 1 - x)^2 \right ]
\label{APsf;b}
\cr
               P^{gq} & = 
                        \frac{4}{3}
                \left [
                        \frac{1 +  (1 - x)^2}{x}
                \right ]
\label{APsf;c}
\cr
               P^{gg} & = 6
                \left [ 
                        \frac{1 - x}{x} 
                        +  x (1  - x) +
                            \frac{x}{(1 - x)_+}
                            +  \frac{11}{12} \delta ( 1 - x)
                \right ]
\cr
               & \qquad -\frac{n_f}{3} 
                        \delta ( 1  -   x)
\label{APsf;d}
\end{array}
$$

Here the gluon distribution $G(x,  Q^2 )$ has been introduced
and $1 / (1 - x)_+$ means
$$
           \int_0^1 dx 
                      \frac{f(x)}{(1 - x)_+} 
=
           \int_0^1 dx\; 
                      \frac{f(x) - f(1)}{(1 - x)} 
\label{gluondis}
$$
Measurement of the structure functions over a large range of $x$ and
$Q^2$ allows both $\alpha_s$ and the parton distributions to be
determined. Notice that $\alpha_s$ {\it and} the gluon distribution
can only be obtained by measuring the $Q^2$ dependence.
 The precision of contemporary experimental data demands that
higher-order corrections also be
included\cite{Curci80}.  The above results are for massless quarks.
Algorithms exist for the inclusion of
nonzero quark masses\cite{Gluck82}. These are particularly important
for neutrino scattering near the charm threshold.  
At low $Q^2$
values, there are also important ``higher-twist'' (HT)
contributions of the form:
$$
                F_i ( x ,  Q^2 ) =
                        F_i^{(LT)}\; (x,  Q^2 )     + 
                        \frac{F_i^{(HT)} (x, Q^2 )}{Q^2}    +  
                         \;\cdots
\label{FixQ2}
$$
\hglue -3pt$\!$
Leading twist (LT) terms are those whose behavior can be predicted using the
parton model, and are related to the parton distribution functions.
Higher-twist corrections depend on matrix elements of higher dimension
operators. These corrections are numerically important only for \hbox{$Q^2 \!<
\!{\cal O}$(few GeV${}^2$)} except for $x$ very close to~1. At very large
values of $x$ corrections proportional to $\log(1-x)$ can become
important\cite{sterman}.

From Eqn.~\ref{AltarelliParisi;a}, it is clear
that a
nonsinglet structure function offers in principle the most
precise test of the theory, since the $Q^2$ evolution is
independent of the unmeasured gluon distribution.  The CCFR collaboration fit
to the Gross-Llewellyn Smith sum rule\cite{gross} which is known to order
$\alpha_s^3$\cite{kataev}\cite{larinj} (Estimates of the order $\alpha_s^4$
term are available\cite{kataev1})
$$
\begin{array}{ll}
        &       \int_0^1 dx \left\{ F_3^{\overline{\nu}p} (  x,  Q^2 )+
F_3^{{\nu}p} ( x ,  Q^2 )\right\} =\cr
        &~~~~           3\left[ 1-\frac{\alpha_s}{\pi}
(1+3.58\frac{\alpha_s}{\pi}
+ 19.0(\frac{\alpha_s}{\pi})^2)\right]-\Delta{HT} \;
\label{lls}
\end{array}
$$
where the higher-twist contribution $\Delta{HT}$ is estimated to be 
$(0.09\pm 0.045)/Q^2 $ in \cite{kataev}\cite{braun} and to be somewhat  
smaller by \cite{dasgupta}.
The CCFR collaboration\cite{ccfrr98}, combines their data with that from 
other experiments \cite{f3rest} and gives 
$\alpha_s~(\sqrt{3} ~\hbox{GeV})=0.28\pm
 0.035 ~(\hbox{expt.}) \pm
0.05 ~(\hbox{sys}) ^{+0.035}_{-0.03}~(\hbox{theory})$. 
The error from higher-twist terms (assumed to be $\Delta{HT}=0.05\pm 0.05$)
dominates the theoretical
error.
If the higher twist result of \cite{dasgupta} is used, the central value 
increases to 0.31 in agreement with the fit of \cite{parente}. This value 
extrapolates to  $\alpha_s~(M_Z)=0.118\pm 0.011$.

Measurements
involving singlet-dominated structure functions, such as $F_2$,
result  in correlated measurements of $\alpha_s$ and the gluon structure function. A full next to leading order fit combining data from 
SLAC\cite{Whitlow}, BCDMS\cite{Benvenuti}, E665\cite{adams}
  and HERA\cite{hera-dis}has been performed \cite{santiago}. 
These authors extend the analysis to next to next to leading order (NNLO). In this case
 the full theoretical calculation is not available as not all the three-loop 
anomalous dimensions are known; their analysis uses moments
 of structure functions\footnote{The moments are defined by $M_n=\int_0^1x^nF(x,Q^2)dx$.} and is restricted to those moments
 where the full calculation is available 
\cite{Curci80,moments,parente}. The NNLO result is 
$\alpha_s~(M_Z)=0.1172\pm
 0.0017 ~(\hbox{expt.}) \pm
0.0017 ~(\hbox{sys})$. 
Here the first error is
 a  combination of
 statistical and systematic experimental errors, and the second error is due
 to the uncertainties,
quark masses, higher twist  and target mass corrections, and errors from the 
gluon distribution. If only a next to leading order fit is performed then the
value decreases to $\alpha_s~(M_Z)=0.116$ indicating that the theoretical 
results are stable. No error is included from the choice of $\mu$;
$\mu=Q$ is assumed. We use a total error of $\pm 0.0045$ to take into
account an estimate of the scale uncertainty.
This result is consistent with earlier determinations
\cite{Bazizi91},
\cite{Virchaux92}, and \cite{Quintas93}.

The spin-dependent structure functions, measured in polarized lepton nucleon 
scattering,
 can also be used  to  
determine $\alpha_s$. 
The spin structure functions   $G_1$  and $G_2$ are 
 defined in terms of the asymmetry
in polarized lepton nucleon scattering
\begin{equation}
 a(x,y)=\frac{d\sigma^{e N}_p}{dxdy}-\frac{d\sigma^{e N}_{ap}}{dxdy}
\end{equation}
where the subscript $p$ ($ap$) refers to
 the state where the nucleon spin is parallel (anti-parallel) to 
its direction of motion in the center of mass frame 
of the lepton-nucleon  system. In both cases the lepton has
its spin aligned along its direction of motion.
\begin{equation}
\begin{array}{ll}
a(x,y)=&\dsp\frac{8\pi \alpha_{em}^2 y}{MQ^2}
((1-2/y^2+2x^2y^2M^2/Q^2)G_1(x,Q^2)\\
&+4x^2M^2G_2(x,Q^2)/Q^2)
\end{array}
\label{asym}
\end{equation}
The $Q^2$ evolution of the spin structure functions $G_1(x,Q^2)$ and
$G_2(x,Q^2)$  is similar to that of the unpolarized ones and is known
at next to leading order \cite{Mertig:1996ny}.Here the values of $Q^2\sim 2.5 \hbox { GeV}^2$ 
are small particularly for the E143 data \cite{abeabe}
and higher-twist corrections are 
important.
A fit\cite{smc98} 
using the measured spin dependent structure functions measured by
themselves and by other experiments \cite{spinsf}\cite{abeabe} gives 
$\alpha_s(M_Z)=0.121\pm 0.002
(\hbox{expt.})\pm 0.006(\hbox{theory and syst.})$. Data from HERMES
\cite{hermes} are not included in this fit; they are consistent with
the older data.

 $\alpha_s$ can also be determined from the Bjorken sum
 rule\cite{bjsum}.
\begin{equation}
S_{{\rm Bj}}
=\int_0^1 dx \left( G_1^{p}-G_1^{n}\right)
=\frac{1}{6}a_3
\end{equation}
  At lowest order in QCD
$a_3 = g_A = \frac{G_V}{G_A}=1.2573\pm 0.0028$. 
A fit gives \cite{altbj}
$\alpha_s(M_Z)=0.118^{+0.010}_{-0.024}$;
a significant contribution to the error being due to the extrapolation
into the (unmeasured) small $x$ region. 
Theoretically, the sum rule is preferable
as the perturbative QCD result  is known to higher order, and these 
terms are important
at the low $Q^2$ involved.
It has been shown that
the theoretical errors associated with the choice of scale 
are considerably reduced
by the use of Pade approximants\cite{ellisbj} which results in
$\alpha_s(1.7~\hbox{GeV})=0.328\pm 0.03 (\hbox{expt.})
\pm 0.025 (\hbox{theory})$ corresponding to
$\alpha_s(M_Z)=0.116^{+0.003}_{-0.005}(\hbox{expt.})\pm 0.003(\hbox{theory})$.
No error is included from the extrapolation into the region of $x$
 that is unmeasured.
Should  data become available at smaller values of
$x$ so that this extrapolation could be more tightly constrained, the sum
rule method could provide a better  determination of $\alpha_s$ than
that  from the spin
structure functions themselves.

At very small values of $x$ and $Q^2$, both the $x$ and $Q^2$ 
dependence of the structure functions
is predicted by perturbative QCD\cite{bfkl}. 
Here terms to all orders in $\alpha_s\ln(1/x)$
are summed.
The data from HERA\cite{hera-dis}  on $F_2^{ep}(x,Q^2) $ 
can be fitted 
to this form\cite{ball}, including the NLO terms which are required to fix
the $Q^2$ scale. 
The data are dominated by $4 \hbox{ GeV}^2<Q^2<100 \hbox{ GeV}^2$.
The fit\cite{Ball96} using H1 data\cite{h194} gives 
$\alpha_s(M_Z)= 0.122\pm 0.004 ~(\hbox{expt.}) \pm 0.009 ~(\hbox{theory})$.
(The theoretical error is taken from \cite{ball}.)
 The dominant
part of the theoretical error is from the scale dependence;
errors from terms that are suppressed by $1/\log(1/x)$ in the quark
sector are included\cite{catha} while those from the gluon sector are
not.

Typically, $\Lambda$ is extracted from the 
deep inelastic scattering data by parameterizing
the parton densities in a simple analytic way at some $Q_0^2 $,
evolving to higher $Q^2$ using the next-to-leading-order
evolution equations, and fitting globally to the
measured structure functions.
Thus, an important by-product of such studies is the extraction
of parton densities at a fixed-reference value of $Q_0^2 $.
These can then be evolved in $Q^2$ and used as input for
phenomenological studies in hadron-hadron collisions (see
below). These densities will have errors associated with the 
that value of $\alpha_s$.  A next-to-leading order fit
must be used if the process being calculated is 
known to next-to-leading order
in QCD perturbation theory. In such a case, there is an additional scheme
dependence; this scheme 
dependence is reflected in the ${\cal O}(\alpha_s)$ corrections
that appear in the relations between the structure functions and the
quark distribution functions. There are two common schemes: a deep-inelastic
scheme where there are no order $\alpha_s$ corrections in the formula for
$F_2(x,Q^2)$ and the minimal subtraction scheme. 
It is important when these next-to-leading
order fits are used in other processes (see below), that 
the same scheme is used in
the calculation of the partonic rates. 
Most current sets of parton 
distributions are obtained using fits to all relevant data 
\cite{partonfits}. 
In particular, data from purely hadronic initial states 
are used as they can 
provide important constraints on the gluon distributions.

\subsubsection{PHOTON STRUCTURE FUNCTIONS}

Experiments in $e^+e^-$ collisions can  be used to study
photon-photon
 interactions and to 
 measure the structure function of a 
photon\cite{witten77}, by selecting events of the 
type $e^+e^-\to e^+e^- + {\rm hadrons} $ which 
proceeds via two photon scattering. If events are selected where
 one of the photons is almost on mass shell and the other has a 
large invariant mass $Q$, then the latter probes the photon structure 
function at scale $Q$; the process is analogous to deep inelastic 
scattering where a highly virtual photon is used to probe the proton structure.
The $Q^2$ variation of this structure function follows that shown above
(see Eq \ref{AltarelliParisi;a}).

A  review of the data can be found in \cite{but99}.
Data have become available from 
LEP\cite{ackerstaff97}
 and from TRISTAN\cite{muramatsu94}\cite{sahu95} 
which extend the range of $Q^2$ to of order 300 GeV$^2$  and $x$ as low as 
$2\times 10^{-3}$and show 
$Q^2$ dependence of the structure function that is consistent with QCD 
expectations. Experiments at HERA can also probe the photon 
structure function by looking at jet production in $\gamma p$ collisions; 
this is 
analogous to the jet production in hadron-hadron collisions which is 
sensitive to hadron structure functions.
The data \cite{h1gamma} are consistent with theoretical models
 \cite{frixone96}.

 \section{$\alpha_s$ FROM  FRAGMENTATION FUNCTIONS}

Measurements of the 
fragmentation function $d_i(z,E)$, 
the probability that a hadron of type $i$ be produced with energy
$zE$ in $e^+e^-$ collisions at $\sqrt{s}=2E$,
can be used to determine $\alpha_s$. As in the case of scaling violations in
structure functions, QCD predicts only the $E$ dependence in a form
similar to the $Q^2$ dependence of Eq \ref{AltarelliParisi;a}.
Hence, measurements at different energies are needed to extract a value
of $\alpha_s$. Because the QCD evolution mixes the 
fragmentation functions for each quark flavor with the gluon fragmentation
function, it is necessary to determine each of these 
before $\alpha_s$ can be extracted. 
The ALEPH collaboration has used data in the
energy range $\sqrt{s}=22$ GeV
 to $\sqrt{s}=91$ GeV. A flavor tag is used to 
discriminate between different quark species,
and the longitudinal and transverse 
cross sections are used to extract the 
gluon fragmentation function\cite{webber95}.
The result obtained is 
$\alpha_s(M_Z)=0.126 \pm 0.007 ~(\hbox{expt.}) 
\pm 0.006 ~(\hbox{theory})$\cite{alephfrag}. 
The theory error is due mainly to
the choice of scale at which $\alpha_s$ is evaluated.
The OPAL collaboration\cite{opalfrag}
has also extracted the separate fragmentation functions.
DELPHI\cite{delphifrag} has performed a similar
analysis using data from other experiments at center of mass energies
between 14 and 91 GeV
with the result 
$\alpha_s(M_Z)=0.124\pm 0.007 \pm 0.009 ~(\hbox{theory})$. The 
larger theoretical 
error is because the value of $\mu$ was allowed to vary between
$0.5\sqrt{s}$ and $2\sqrt{s}$.
These results can be combined to give
 $\alpha_s(M_Z)=0.125\pm 0.005 \pm 0.008 ~(\hbox{theory})$.

\section{$\alpha_s$ FROM EVENT SHAPES AND JET COUNTING}

An alternative method of determining $\alpha_s$ in  $e^+ e^-$
annihilation involves measuring the
the topology of the hadronic final states.
There are many possible
choices of inclusive event shape variables:  thrust\cite{Farhi77}, energy-energy
correlations\cite{Basham78},
average jet mass, {\it etc.}.  These quantities must be infrared safe, 
which means that they are insensitive to the low energy properties of
QCD 
and can therefore be reliably 
calculated in perturbation theory. For example, the thrust distribution is
 defined by  
\begin{equation}T=max(\sum_i
\abs{\overline{p_i}\mdot\overline{n}}/\sum_i\abs{\overline{p_i}})
,\end{equation} where the
sum runs over all hadrons in the final state and the unit vector $\overline{n}$
is varied.  At lowest order in QCD the process $e^+e^- \to
q\overline{q}$ results in a final state with back to back quarks {\it
  i.e.}  ``pencil-like'' event with  $T=1$.  Alternatively, the event
can be  divided by a plane 
normal to the thrust axis and the invariant mass of the particles in
the two hemispheres is computed, the larger (smaller)  of these is
$M_h$ ($M_l$). At lowest order in QCD $M_h= M_l=0$.

The observed final state consists of
hadrons rather than the quarks and gluons of perturbation theory. The
hadronization of the partonic final state has an energy scale of order 
$\Lambda$. The resulting hadrons acquire momentum components
perpendicular to the original quark direction of order $\Lambda$.
This effect induces corrections to the shape variables of order
$\Lambda/\sqrt{s}$. A model is needed 
to describe the detailed evolution
of a partonic final state into one involving hadrons, so that detector
corrections can be applied. Furthermore if
the QCD matrix elements are combined with a  
parton-fragmentation model, 
this model can then be used to correct the data for a
direct comparison with the perturbative QCD calculation.
The different hadronization models that are 
used\cite{Andersson83}
  model the 
dynamics that are controlled by non-perturbative QCD effects
which we cannot yet calculate.  The fragmentation parameters of these 
Monte Carlo simulations are
tuned to get agreement with the observed data.
 The differences between these models can be used to estimate
 systematic errors.

In addition to using a shape variable, one can perform a jet counting
experiment. At order $\alpha_s$ the partonic final state
$q\overline{q}g$ appears which can manifest itself as a three-jet
final state after hadronization. Every higher order produce a higher
jet multiplicity and
measuring quantities that are sensitive to the
relative rates of two-,  three-, and four-jet events can lead to a
determination of $\alpha_s$.
There are theoretical ambiguities
in the way that particles are combined to form jets. 
Quarks and gluons are massless,
whereas the observed hadrons are not,
 so that the massive jets that result from
combining them  cannot be compared directly to the
massless jets of perturbative QCD.

The jet-counting algorithm, originally
introduced by the JADE collaboration\cite{Bethke88}, has
been used by many other groups.
Here, particles of momenta $p_i$ and $p_j$ are combined into a
 pseudo-particle of momentum $p_i+p_j$ if the invariant mass of the pair is
 less than $y_0\sqrt{s}$.
 The process is then iterated until no more pairs of
 particles or pseudo-particles remain.
 The remaining number of pseudo-particles is then defined to
 be the number of jets in the event, and 
can be compared to the perturbative QCD prediction which depends on $y_0$.
The Durham algorithm is slightly different: in computing the mass
of a pair of partons, it uses 
$M^2=2\hbox{min}(E_1^2,E_2^2)(1-\cos\theta_{ij})$ for partons of energies
$E_i$ and $E_j$ separated by angle $\theta_{ij}$\cite{durham}.
Different recombination
schemes have been tried,
 for example combining 3-momenta and then rescaling the
energy of the cluster so that it remains  massless.
These varying schemes result in the same data giving slightly different 
values\cite{Akrawy91}\cite{fixedAbe95} of
$\alpha_s$.  These differences can be used to estimate a
systematic error. However, such an error may be conservative as it is
not based on a systematic approximation.

The starting point for all these quantities is the
multijet cross section. For example, at order $\alpha_s$, for  the process
$e^+ e^- \to qqg$:\cite{e-jet}
$$
                \frac{1}{\sigma}\;
                \frac{d^2 \sigma}{dx_1 dx_2}
=
                \frac{ 2 \alpha_s}{ 3 \pi}\;
                \frac{x_1^2  +  x_2^2}{(1 - x_1 )(1 -  x_2)}
\label{1oversigma}
$$
                where $x_i = \frac{2E_i}{\sqrt s}$are the center-of-mass energy fractions
of the final-state (massless) quarks.  
The order $\alpha_s^2$ corrections to this 
process have been computed, as well as the 4-jet final states such as  $e^+ e^- \to qqgg$\cite{Ellis80}. A distribution in a
``three-jet'' variable, such as those listed above, is obtained by
integrating this differential cross section over an appropriate phase
space region for a fixed value of the variable. Thus $<1-T> \sim
\alpha_s$,
$<M_h^2>/s\sim \alpha_s$ and $<M_l^2>/s\sim \alpha_s^2$.


The result of this integration depends explicitly on $\alpha_s$ but
scale $\mu$ at which $\alpha_s(\mu)$ is to be evaluated is not clear. In
the case of jet counting,
the invariant
mass of a typical jet (or $\sqrt{sy_{0}}$)
is probably a more appropriate choice
than the $e^+e^-$ center-of-mass energy. While there is
no justification for doing so, if the value of $\mu$ is allowed to float in
the fit to the data, the fit improves and  the data tend to prefer
 values of order
$\sqrt{s}/10$ GeV  for some variables 
\cite{fixedAbe95}\cite{opal};
 the
exact value depends on the variable that is fitted. Typically
experiments assign a systematic error from the choice of $\mu$ by
varying it by a factor of 2 around the value determined by the
fit. The choice of this factor is arbitrary

Estimates for the non-perturbative corrections to
$<1-T>$ have been made \cite{dok95} using an operator product
expansion. 
\begin{equation}
<1-T>=A\frac{\alpha_s(\mu)}{2\pi}+B(\frac{\alpha_s(\mu)}{2\pi})^2+C\frac{\alpha_0}{\sqrt{s}}
\label{eq:thrust}
\end{equation}
where A and B known quantities \cite{Ellis80}, $\mu$ is the
renormalization scale and $\alpha_0$ is the non-perturbative parameter
 (the matrix element of an appropriate operator)
to be determined from experiment. Note that the corrections are only
suppressed by $\sqrt{s}$. This  provides an alternative to the
use of
 hadronization
models for estimating these non-perturbative corrections.  
 The  DELPHI collaboration
 \cite{delphi99} uses  data below the $Z$ mass from many experiments
 and Eq.~\ref{eq:thrust} to 
determine
$\alpha_s(M_Z)=0.119\pm 0.006$, the error being dominated 
by the choice of scale. The values of $\alpha_s$ and the
non-perturbative parameter $\alpha_0$ are  also determined by a fit 
to using the  variable $<M_h^2>/s$. While 
the extracted values of $\alpha_s(M_Z)$ 
are consistent with each other, the values of $\alpha_0$ 
are not. The analysis is useful 
as one can directly determine the size of the $1/E$ corrections; 
they are approximately 20\% (50\%) of the perturbative result 
at $\sqrt{s}=91 (11)$ GeV. Even at $\sqrt{s}=91$ GeV the omission of
these perturbative terms will cause a shift on the extracted value of
$\alpha_s$ of $\sim 0.05$ which is much larger than the quoted
experimental errors.

The perturbative QCD formulae can break down in special kinematical
configurations. For example, the first term in Eq. \ref{eq:thrust}
contains a term 
of the type
$\alpha_s\ln^2(1-T)$. The higher orders in the perturbation
 expansion contain terms
of order $\alpha_s^n\ln^m(1-T)$. For $T\sim 1$ (the region populated by 2-jet
events), the perturbation expansion in $\alpha_s$ is
unreliable. The terms with $n\le m$ 
can be summed to all orders in $\alpha_s$\cite{catani}.
 If the jet recombination methods are used, higher-order terms involve
$\alpha_s^n\ln^m(y_0)$, these too can be resummed\cite{webber93}.
 The resummed results give better agreement with the data at
 large values of $T$.
Some caution should be exercised in using these resummed results because of the
possibility of overcounting; the showering Monte Carlos that are used for the
fragmentation corrections  also generate some of these leading-log corrections.
Different schemes for combining the order $\alpha_s^2$ and
the resummations are available\cite{matching}.
These different schemes result in shifts in $\alpha_s(M_Z)$ of
order $\pm 0.002$. 
The use of the resummed results 
improves the agreement between the data and the theory.

Studies on event shapes have been undertaken at lower energies
at 
TRISTAN, PEP/PETRA, and  CLEO. 
A combined result from various shape parameters by the TOPAZ
collaboration gives $\alpha_s(58~ \hbox{GeV})=0.125\pm 0.009$, 
using the fixed order QCD
result, and $\alpha_s(58~ \hbox{GeV})=0.132\pm 0.008$ 
(corresponding to $\alpha_s(M_Z)=0.123\pm 0.007$) where the error is
dominated by scale and fragmentation uncertainties.
The CLEO collaboration fits
 to the order $\alpha_s^2$ results for the two 
jet fraction at $\sqrt{s}=10.53$ GeV, and obtains
$\alpha_s(10.93)=0.164\pm 
0.004~(\hbox{expt.})\pm 0.014~(\hbox{theory})$\cite{cleoshape}. 
The dominant systematic error arises from the choice
of scale ($\mu$), and  is determined from the range of 
$\alpha_s$ that results from  fit with $\mu=10.53$ GeV, and a fit 
where $\mu$ is allowed to vary to get the lowest $\chi^2$. 
The latter results in $\mu=1.2$ GeV. Since the quoted result
corresponds to $\alpha_s(1.2)=0.35$,
 it is by no means clear that the perturbative 
QCD expression is reliable and the  resulting error
should, therefore, be treated with caution. 
A fit to  many different variables as is done in the
LEP/SLC analyses would give added  confidence to the quoted error.

Recently studies have been carried out at 
energies between $\sim$130~GeV\cite{lep130} and $\sim$189~GeV \cite{lep190}. 
These can be combined to 
give $\alpha_s(130~\hbox{GeV})=0.114\pm 0.008$ and  
  $\alpha_s(189~\hbox{GeV})=0.1104\pm 0.005$.
The dominant errors are theoretical and systematic and, as most of 
these are in common at the different energies, these data, those at
 the $Z$ resonance and lower energy provide very  
clear confirmation of the expected decrease
in $\alpha_s$ as the energy is increased.

A combined analysis of the data between 35 and 189 GeV using data
from OPAL and JADE\cite{opaljade} using a large set of shape variables 
shows excellent agreement with $\alpha_s(M_Z)=0.1187^{+0.0034}_{-0.0019}$.
A comparison of this result with those  at the $Z$ resonance 
from SLD\cite{fixedAbe95},
OPAL\cite{opalshape}, L3\cite{l3shape}, ALEPH\cite{alephshape}, and
DELPHI\cite{delphishape}, indicates that they are all consistent with
this value. The experimental errors are smaller than the theoretical
ones arising from choice of scale $\mu$ and modeling of non-perturbative
effects, which are common to
all of the experiments. The SLD collaboration \cite{fixedAbe95}
determines the allowed range of $\mu$ by allowing any value that is
consistent with the fit. This leads to a larger error ($\sim 0.0056$) than that
obtained by DELPHI \cite{delphishape} who vary  $\mu$ by a factor of 2
around the best fit value and obtain $\pm 0.0008$. We elect to use a
more conservative average of  $\alpha_s(M_Z)=0.119\pm 0.005$.

 At lowest order
in $\alpha_s$, the $ep\to e X $ scattering 
process produces a final state of (1+1) jets,
 one from the proton fragment and 
the other from the quark knocked out 
by the underlying process $e+\hbox{quark}\to e+\hbox{quark}$. 
At next order in $\alpha_s$, a gluon can be radiated, and
hence a (2+1) jet final state produced. 
By comparing the rates for these (1+1) and (2+1) jet 
processes, a value of $\alpha_s$ can be obtained. 
A NLO QCD calculation is available\cite{graudenz}.
The basic methodology is similar to that used in the 
jet counting experiments in $e^+e^-$ annihilation discussed above.
Unlike those  measurements, the ones in 
$ep$ scattering are not at a fixed value
of $Q^2$. In addition to the 
systematic errors associated with the jet definitions, 
there are additional ones since the structure functions enter
into the rate calculations. 
Results from H1\cite{h195} and 
ZEUS\cite{zeus95} can be combined to give 
$\alpha_s(M_Z)=0.118 \pm 0.0015 ~(\hbox{stat.}) \pm 0.009
~(\hbox{syst.})$. The contributions to the 
systematic errors from experimental effects 
(mainly the hadronic energy scale of the calorimeter)
 are comparable to  
 the theoretical ones arising 
from scale choice, structure functions, and jet
definitions. 
The theoretical  errors are common to the two 
measurements; therefore, we have 
not reduced the systematic error after forming the
average.

\section{$\alpha_s$ FROM $\tau$ DECAY}

The coupling constant $\alpha_s$ can be determined from an analysis of hadronic
$\tau$ decays~\cite{pich,Braatennar,braaten}. The quantity that will be used
is the ratio
\begin{equation}
R_\tau = { \Gamma(\tau \to \nu_\tau + {\rm hadrons} + (\gamma)) \over
\Gamma(\tau \to \nu_\tau e \bar \nu_e + (\gamma))},
\end{equation}
where $(\gamma)$ represents possible electromagnetic radiation, or lepton pairs.
In the absence of radiative corrections, the ratio $R_\tau$ is
\begin{equation}
R_\tau = 3 \left(  \abs {V_{ud}}^2 + \abs{V_{us}}^2 \right) \approx 3,
\end{equation}
where $3$ is the number of colors. The experimental value $R_\tau = 3.61 \pm
0.05$ is close to three, which is experimental evidence for the existence of
three colors in QCD. The deviation of $R_\tau$ from three is used to extract
$\alpha_s$.

\begin{figure}
\epsfxsize=10cm
\hfil\epsfbox{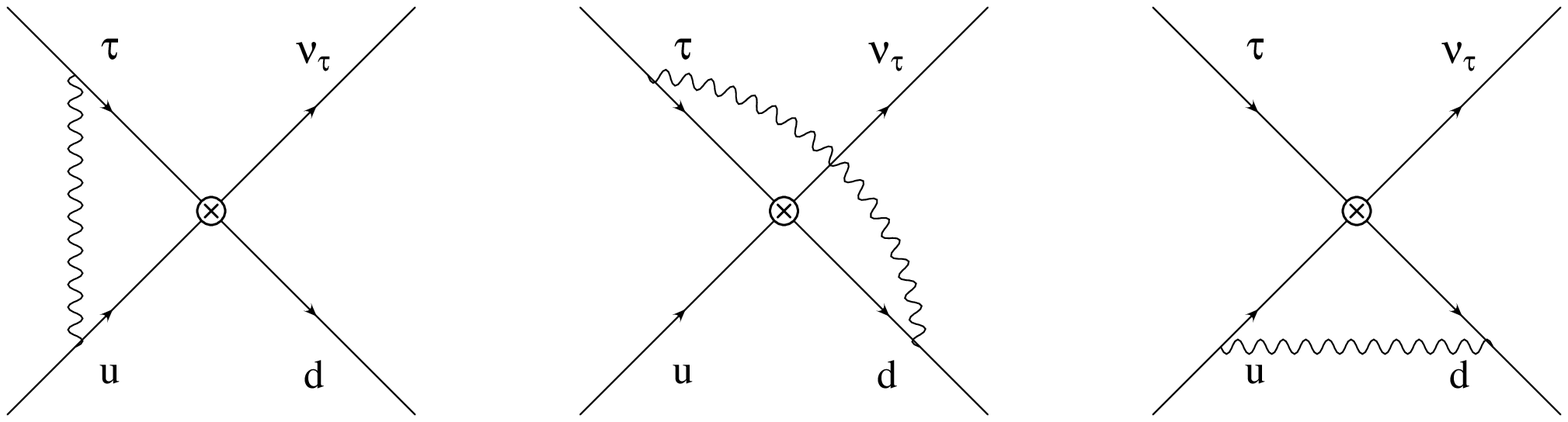}\hfill
\caption{Electromagnetic corrections to the $\tau$ decay vertex.
\label{fig:tauren}}
\end{figure}

The weak decay Lagrangian for non-leptonic $\tau$ decay
is
\begin{eqnarray}\label{9}
L = -{4 G_F \over \sqrt 2} C_\tau(\mu)
\left[ V_{ud}^* \bar \nu_\tau \gamma^\mu P_L \tau
\bar d \gamma_\mu P_L u + 
V_{us}^* \bar \nu_\tau \gamma^\mu P_L \tau \bar s \gamma_\mu P_L u\right],
\end{eqnarray}
where $V_{us}$ and $V_{ud}$ are the CKM mixing angles. The Lagrangian 
Eq.~(\ref{9}) is obtained at the scale $\mu=M_W$ by integrating out the $W$
boson to generate a local four-Fermion operator in the effective theory below
$M_W$, and $C_\tau=1$ at $\mu=M_W$. The typical momentum transfer in $\tau$ decays is of order $m_\tau$, so
it is necessary to scale the Lagrangian Eq.~(\ref{9}) from $\mu=M_W$ to
$\mu=m_\tau$. 
Electromagnetic interactions renormalize the Lagrangian. At one-loop, the
renormalization from graphs shown in Fig.~(\ref{fig:tauren})  
produce a multiplicative renormalization of the Lagrangian, and give
\begin{equation}\label{10}
C_\tau(m_\tau)=1 + {\alpha_{\rm em} \over \pi} \ln {M_W \over m_\tau} \approx
1.009.
\end{equation}

The $\tau$ decay amplitude $\tau \to \nu_\tau X$, where $X$ is the final
hadronic state, can be written as
\begin{equation}
A = -i{4 G_F \over \sqrt 2} C_\tau(\mu) \bar u (p_{\nu}) 
\gamma^\mu P_L u(p_\tau) \left[V_{ud}^* \me{X}{\bar d \gamma_\mu P_L u }{0}
+ V_{us}^*\me{X}{\bar s \gamma_\mu P_L u }{0} \right].
\end{equation}
Squaring the amplitude, and computing the decay rate gives
\begin{eqnarray}
\Gamma&=&4{G_F^2 \abs{C_\tau}^2 \abs{V_{ud}}^2 \over m_\tau}
\sum_X \int {d^3 p_\nu \over (2\pi)^3 2 E_\nu} \left[
p_{\nu}^\mu p_\tau^\nu +
p_{\nu}^\nu p_\tau^\mu - p_{\nu} \cdot p_\tau g^{\mu \nu} -i \epsilon^{\mu
\nu}{}_{\alpha \beta} p_{\nu}^\alpha p_\tau^\beta \right]\nonumber \\
&&\times (2\pi)^4 \delta^4(p_\tau-p_\nu-p_X)
\me{0}{\bar u \gamma_\nu P_L d }{X}\me{X}{\bar d \gamma_\mu P_L u }{0},
\label{13a}
\end{eqnarray}
where we have retained only the $V_{ud}$ term for simplicity. The sum on $X$ is
symbolic for the sum over all final states, including phase space factors. The
$\delta$ function can be written as
\begin{equation}\label{12}
\delta^4(p_\tau-p_\nu-p_X) = \int d^4 q
\delta^4(p_\tau-p_\nu-q)\delta^4(q-p_X),
\end{equation}
and the sum on $X$ can be written as
\begin{equation}
\sum_X (2\pi)^4 \delta^4(q-p_X)
\me{0}{\bar u \gamma_\nu P_L d }{X}\me{X}{\bar d \gamma_\mu P_L u }{0}
= W_{\nu \mu}^{(ud)}(q).
\end{equation}
The tensor $W_{\mu \nu}$ is related to another quantity $\Pi_{\mu \nu}$, defined
by
\begin{equation}
\Pi_{\mu\nu}^{(ud)}(q)= -i \int {d^4 q }e^{-i q \cdot x}
\me{0}{T( j_\nu^\dagger(x) j_\mu(0) }{0},
\end{equation}
where $j^\mu=\bar d \gamma_\mu P_L u$. Inserting a complete set of states in the
time-ordered product, one finds that
\begin{equation}
W_{\mu\nu}^{(ud)}= 2\ \im \Pi_{\mu \nu}^{(ud)}
\end{equation}

The tensor $\Pi_{\mu\nu}$ depends on the only variable, $q$, and must have the form
\begin{equation}
\Pi_{\mu \nu}^{(ud)}(q) = \left(-q^2 g_{\mu \nu} +  q_\mu q_\nu\right)
\Pi_T^{(ud)}(q^2) + q_\mu q_\nu \Pi_L^{(ud)}(q^2),
\end{equation}
by Lorentz invariance. The tensor $W_{\mu \nu}$ is then given by
\begin{equation}\label{11}
W_{\mu \nu}^{(ud)}(q) = \left(-q^2 g_{\mu \nu} + 
 q_\mu q_\nu\right)\Omega_T^{(ud)}(q^2) + q_\mu q_\nu \Omega_L^{(ud)}(q^2),
\end{equation}
where
\begin{equation}
\Omega_L^{(ud)} = 2\ \im \Pi_L^{(ud)},\qquad \Omega_T^{(ud)} = 2\ \im \Pi_T^{(ud)}.
\end{equation}
If the light quark mass difference $m_d-m_u$ and $m_s-m_u$ are neglected, 
the hadronic currents $\bar d \gamma_u P_L u$ and $\bar s \gamma_u P_L u$ are
conserved. This implies that $q^\mu \Pi_{\mu\nu}=0$, so that $\Pi_L(q^2)=0$.
Inserting Eq.~(\ref{12}) and Eq.~(\ref{11}) into Eq.~(\ref{13}) gives
\begin{eqnarray}
\Gamma&=&2{G_F^2 \abs{C_\tau}^2  \over m_\tau}
\int d^4 q \delta(p_\tau-p_\nu - q)
 \int {d^3 p_\nu \over (2\pi)^3 2E_\nu} \nonumber \\
&&(m_\tau^2-q^2)
\left[\Omega_T (m_\tau^2 + 2 q^2 ) + \Omega_L m_\tau^2 \right],
\label{13}
\end{eqnarray}
where 
\begin{equation}
\Omega_{T,L}=\abs{V_{ud}}^2 \Omega_{T,L}^{(ud)} + \abs{V_{us}}^2 \Omega_{T,L}^{(us)}
\end{equation}
and we added back the $V_{us}$ contribution. There is no interference term (at lowest order in
the weak interactions), because the
$u \to d$ and $u \to s$ currents lead to final states with different flavor
quantum numbers, and the strong interactions conserve flavor. 
The hadronic invariant mass
distribution can then be written as
\begin{eqnarray}
{d\Gamma \over ds} &=&{G_F^2 \abs{C_\tau}^2 \over 8 \pi^2 
m_\tau^3}(m_\tau^2-s)^2
\left[\Omega_T(s) (m_\tau^2 + 2 s) + \Omega_L(s) m_\tau^2 \right]
\label{15}
\end{eqnarray}
The ratio of the
hadronic to leptonic decay rate of the $\tau$ is given by~\cite{Braatennar}
\begin{equation}\label{20}
R_\tau = 6\pi\abs{C_\tau}^2\int_0^{m_\tau^2} {ds \over m_\tau^2} \left(1-{s \over m_\tau^2}
\right)^2 \left[\Omega_T(s)\left(1+{2 s \over m_\tau^2}
\right)+\Omega_L(s) \right]
\end{equation}

\begin{figure}
\epsfxsize=5cm
\hfil\epsfbox{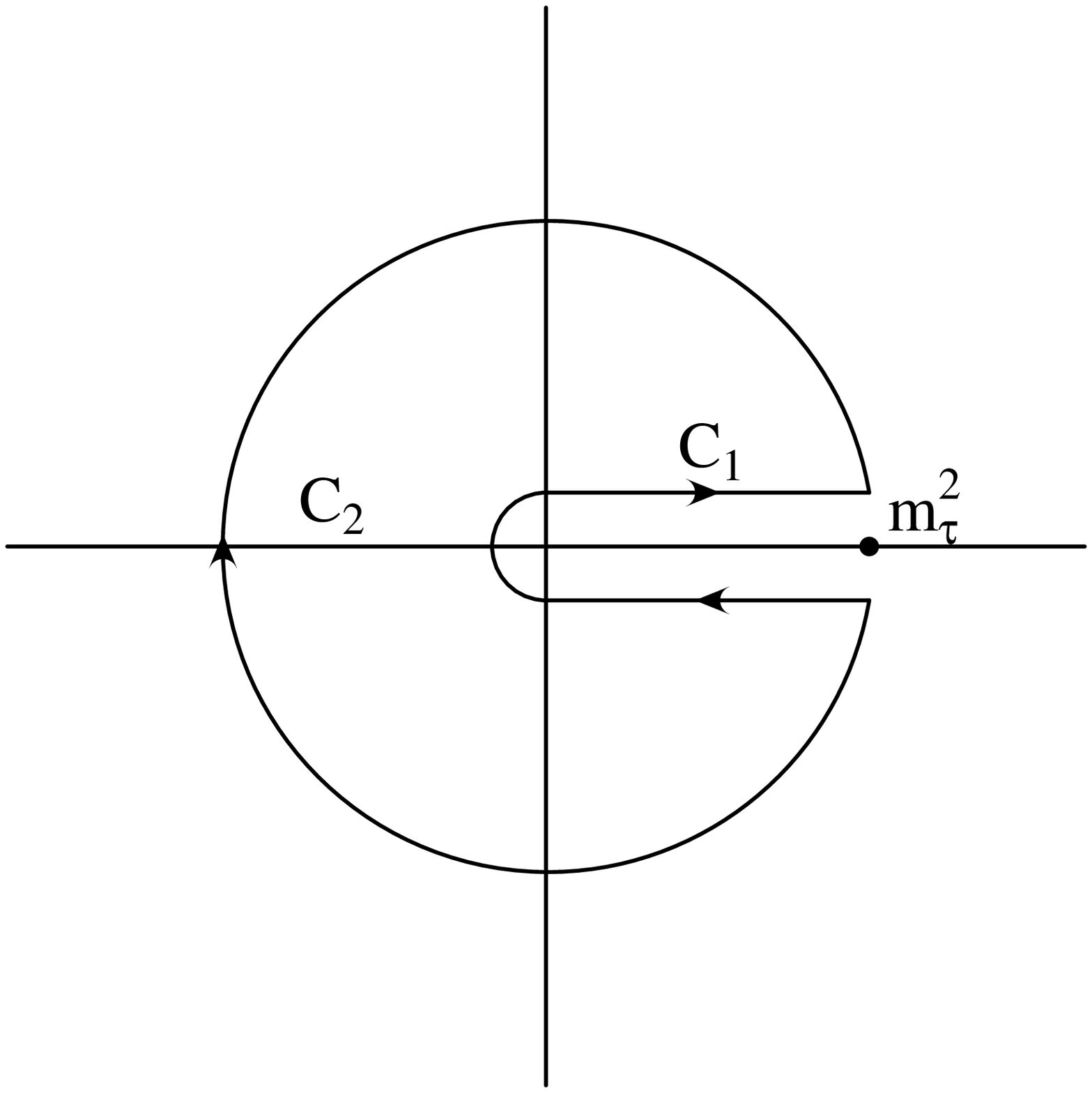}\hfill
\caption{Integration contours $C_1$ and $C_2$ in the complex $s$ plane.
\label{fig:taucontour}}
\end{figure}

The hadronic tensors $\Pi_{L,T}(s)$ are analytic in the complex $s$ plane,
except for a branch cut along the positive real axis. The discontinuity across
the cut is $\Omega_{L,T}(s)$, and is the cross-section for the currents to
create hadrons. Clearly, the hadron production rate is sensitive to
non-perturbative effects, and can not be computed reliably. Far away from the
physical cut, there are no infrared singularities, and QCD perturbation theory
is valid. One can rewrite the integral Eq.~(\ref{20}) as
\begin{equation}\label{21}
R_\tau = 6\pi i\int_C {ds \over m_\tau^2} \left(1-{s \over m_\tau^2}
\right)^2 \left[\Pi_T(s)\left(1+{2 s \over m_\tau^2}
\right)+\Pi_L(s) \right],
\end{equation}
where $C=C_1$ is the contour shown in Fig.~(\ref{fig:taucontour}).
The difference of $\Pi_{L,T}$ above and below the cut is $\Omega_{L,T}$, so this
gives back Eq.~(\ref{20}). Since $\Pi_{L,T}$ have no singularities in the
complex $s$ plane other than the branch cut, one can deform the contour $C_1$ to
the contour $C_2$, and use Eq.~(\ref{21}) with $C=C_2$. The advantage of using
Eq.~(\ref{21}) with $C=C_2$ rather than $C=C_1$ is that one needs to know
$\Pi_{L,T}$ far away from the cut for most of the integration contour. The
contour approaches the cut at $s=m_\tau^2$, but at this point, the integrand
vanishes as $(s-m_\tau)^2$, so the contribution of the region near $s=m_\tau^2$
to the total integral is suppressed~\cite{Braatennar}.

$\Pi(s)$ can be computed in perturbation theory using an operator product
expansion, which is valid away from the physical cut. The perturbation theory
result for $\Pi(s)$ is then substituted in Eq.~(\ref{21}). In practice, the
calculation can be simplified by using the perturbation theory value for
$\Omega(s)$ in Eq.~(\ref{20}). We have argued above that perturbation theory is
not valid for $\Omega(s)$. Nevertheless, using the perturbation theory value for
$\Omega(s)$ in Eq.~(\ref{20}) is justified because $\Pi(s)$ in perturbation
theory has the same analytic structure in QCD. Thus using the perturbation
theory value of $\Pi(s)$ in Eq.~(\ref{21}) is equivalent to using the
perturbation theory value for $\Omega(s)$ in Eq.~(\ref{20}), even though the
perturbative computation of $\Omega(s)$ is not valid.
 
The OPE for $\Pi(s)$ is
closely related to that for $e^+ e^- \to {\rm hadrons}$, which depends on
the time-ordered product of two electromagnetic currents.
\begin{equation}
\Pi(s) = c_i(\mu,s,\alpha_s(\mu)) \vev{O_i(\mu)},
\end{equation}
where $c_i$ are the coefficient functions, and $O_i$ are the local operators.
Since the contour $C_2$ is a circle of radius $m_\tau^2$ in the complex $s$
plane, one expects that logarithms in the uncalculated higher order corrections
are minimized if one chooses $\mu=m_\tau$. The leading order operator is the
unit operator. In the limit that the light quark masses are neglected, $\Pi_L$
vanishes, and we only need to compute $\Pi_T$, giving
\begin{eqnarray}
2 \pi \Omega_T(s) &=& \abs{C_\tau}^2\left(\abs{V_{ud}}^2 + \abs{V_{us}}^2\right)
\nonumber\\ &&\times \left[
1 + {\alpha_s(\sqrt{s}) \over \pi} + F_3 
\left({\alpha_s(\sqrt{s}) \over \pi}\right)^2 + F_4
\left({\alpha_s(\sqrt{s} )\over \pi}\right)^3 + \ldots  \right].
\end{eqnarray}
The first coefficient is the well-known result that the ratio $\sigma(e^+ e^-
\to {\rm hadrons})/\sigma(e^+ e^- \to q \bar q)$ is
$3(1+\alpha_s(\sqrt{s})/\pi)$. The next two coefficients are
\begin{equation}
F_3 = 1.9857-0.1153n_f,\qquad F_4 = -6.6368-1.2001n_f-0.0052 n_f^2.
\end{equation}
Using the $\beta$-function to write $\alpha_s(\sqrt{s})$ in terms of
$\alpha_s(m_\tau)$, evaluating the $s$ integral, and setting $n_f=3$ 
gives~\cite{Braatennar}
\begin{eqnarray}
R_\tau &=& 3 \abs{C_\tau}^2\left(\abs{V_{ud}}^2 + \abs{V_{us}}^2 \right)
\nonumber\\
&& \times \left[1+
{\alpha_s(m_\tau) \over \pi} + 5.2023 
\left({\alpha_s(m_\tau) \over \pi}\right)^2 + 26.366
\left({\alpha_s(m_\tau) \over \pi}\right)^3 +\ldots \right]\label{50}
\end{eqnarray}
Assuming $\alpha_s(m_\tau)\approx 0.35$, the series in brackets is
$1 + 0.111 + 0.065 + 0.036 + \ldots$, so the terms are still numerically
decreasing till order $\alpha^3$. One can use the value of the last term as an
estimate of the theoretical uncertainty in the perturbative value for the
coefficient of the unit operator. This also tells us that we can neglect
corrections from higher dimension operator that are smaller than about $3\%$.

The $1/m_\tau^2$ corrections to $R_\tau$ arise from the quark mass corrections
to the coefficient of the unit operator in the OPE. At order $m^2/m_\tau^2$, the
currents are no longer conserved, so one needs to compute both $\Pi_L$ and
$\Pi_T$. The only light quark mass contribution of any significance is the
$s$-quark mass correction\cite{Braatennar}, 
\begin{equation}
\delta R_\tau = 3 \abs{V_{us}}^2\left[ -8 {m_s^2 \over m_\tau^2}
 \left(1 + {16 \alpha_s(m_\tau) \over 3
\pi} \right)\right].
\end{equation}
Using $m_s\sim 150$~MeV as an estimate for the $s$-quark mass gives $\delta
R_\tau \sim -0.008$, which is smaller than the error in the perturbation series.

The $1/m_\tau^4$ and  corrections in the OPE arise from the dimension four
operators  $F_{\mu \nu} F^{\mu \nu}$ and $m \bar \psi \psi$, and the
$1/m_\tau^6$ corrections from the four-quark operators $\bar\psi \Gamma \psi
\bar \psi \Gamma \psi$, where $\Gamma$ is some combination of $\gamma$
matrices. An analysis of these corrections, based on model estimates of the 
operator matrix elements indicates that these corrections are smaller than the
uncertainty in the perturbation series\cite{Braatennar}. The size of
non-perturbative corrections can be determined directly from the experimental
data. Instead of considering the integral Eq.~(\ref{20}) that gives the total
hadronic width, one compares the integral of $d\Gamma/ds$ weighted with

$(1-s/m_\tau^2)^k (s/m_\tau^2)^l$ with the corresponding moment of the
experimental data. By studying the moments for different values of $k$ and $l$,
one finds that the non-perturbative corrections are about
3\%~\cite{alephtau,opaltau}, and so are comparable in size to the uncertainty
in the perturbation series.

The experimentally measured quantity is
$R_\tau^{(ud)}=3.484\pm0.024$~\cite{cleotau,alephtau}, the ratio
for $\tau$ to decay into non-strange hadrons to the leptonic decay 
rate. This is
given by Eq.~(\ref{50}), dropping $V_{us}$, and gives
\begin{equation}
\alpha_s(m_\tau)=0.34 \pm 0.03,
\end{equation}
where we have assumed a theoretical uncertainty of 100\% in the
$\alpha^3$ term.
This value corresponds to \begin{equation}
\alpha_s(m_Z)=0.119 \pm 0.003,
\end{equation}

\section{$\alpha_s$ FROM LATTICE GAUGE THEORY COMPUTATIONS}

The strong coupling constant $\alpha_s$ can be determined from lattice gauge
theory calculations of the hadronic spectrum. The basic procedure used is to
choose a definition of $\alpha_s$, and measure its value on the lattice. One
then has to set the scale at which $\alpha_s$ takes on the measured value. The
lattice scale can be normalized using the hadronic spectrum measured on the same
lattice. Finally, one has to convert the lattice definition of $\alpha_s$ to the
value defined in the continuum in a scheme such as \msbar. 

There are several sources of systematic errors that limit the current accuracy
in determining $\alpha_s$. Typically, $\alpha_s$ is determined by determining
the spectrum of heavy quark bound states on the lattice. There are corrections
due to the finite volume and finite lattice spacing $a$. The finite lattice
spacing errors can be reduced by using improved actions, that are accurate to
higher order in $a$. To some extent, one can estimate the error due to finite
volume and finite lattice spacing by repeating the simulation on a larger
lattice. The dominant systematic uncertainty is due to the quenched
approximation, in which light quark loops are neglected. It is difficult to
reliably estimate the systematic errors due to this approximation without doing
a full simulation including dynamical fermions. Simulations with dynamical
fermions are just starting to be done, and in a few years one should have more
reliable estimates of $\alpha_s$. 

There is one important advantage to using a heavy quark system such as the
$\Upsilon$ to determine $\alpha_s$. The leading correction to the $\Upsilon$
energy levels due to the light quark masses is linear in the quark masses, and
can only depend on the flavor singlet combination $m_u+m_d+m_s$. Thus the light
quark mass corrections can be computed to a good approximation using three
light quarks of mass $(m_u+m_d+m_s)/3$. This avoids having to simulate almost
massless dynamical quarks, which is very difficult.

Lattice calculations can also be used to test theoretical calculations, and
determine the regime in which perturbation theory is applicable. In the
quenched approximation, one can study the scale dependence of the coupling
constant on the lattice. This provides a check on the perturbation theory
calculation with $n_f=0$. The result is in remarkable agreement with the
perturbation theory result in the regime where the coupling constant is
weak~\cite{gaugnelli,mlus}.

The Fermilab and SCRI groups use the $S-P$ and $1S-2S$ splittings in the
$\Upsilon$ system to determine $\alpha_s$.
There are some systematic deviations of the calculated numbers from their
experimental values in the quenched approximation ($n_f=0$), which are
dramatically reduced if one includes $n_f=2$ dynamical flavors. The value of
$\alpha_s(M_Z)$ in the \msbar\ scheme is~\cite{Khadra96}
\begin{equation}
\alpha_s(M_Z)=0.1159  \pm  0.0019 \pm 0.0013 \pm 0.0019
\end{equation}
where the first error is due to discretization effects, relativistic
corrections, and statistical errors, the second is due to dynamical fermions,
and the third is from conversion uncertainties.

More recent computations give (in the $n_f=5$ scheme)~\cite{davies97}
\begin{equation}
\alpha_s(M_Z)=0.1174  \pm  0.0024 
\end{equation}
and~\cite{sesam99}
\begin{equation}
\alpha_s(M_Z)=0.1118  \pm  0.0017. 
\end{equation}
An average of these newer values gives $\alpha_s(M_Z)=0.115  \pm  0.004$, where
we have included the difference between the two central values as an estimated
additional systematic error.

\section{$\alpha_s$ FROM HEAVY QUARK SYSTEMS}

\def\bsigma{\mbox{\boldmath $\sigma$}}

\begin{figure}
\epsfxsize=8cm
\hfil\epsfbox{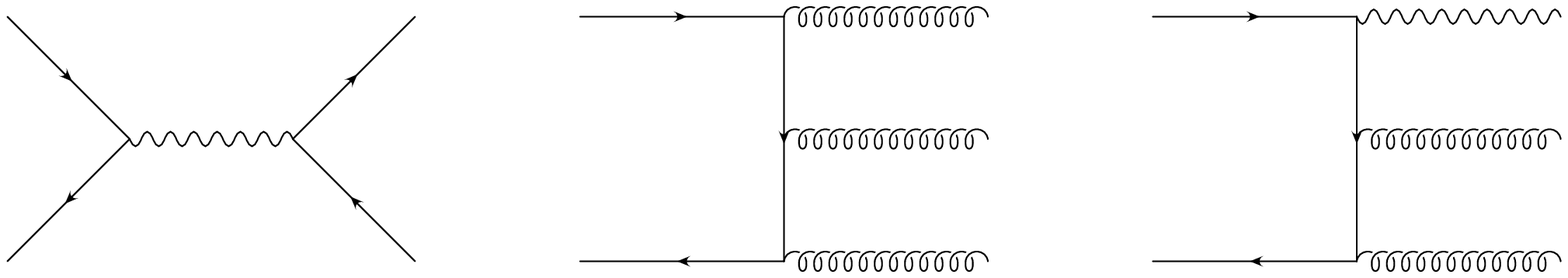}\hfill
\caption{Decay diagrams for $\Upsilon \to \mu^+ \mu^-$, $\Upsilon \to ggg$ and
$\Upsilon \to \gamma gg$.
\label{fig:upsdecay}}
\end{figure}

Heavy quark bound states such as the $\Upsilon$ can also be used to extract a
value for $\alpha_s$. If the bound state is treated using non-relativistic
quantum mechanics, the annihilation decays $\Upsilon \to \mu^+ \mu^-$,
$\Upsilon \to ggg$ and $\Upsilon \to \gamma g g$ can be computed as the product
of the probability to find the quark-antiquark pair at the origin, times the
annihilation rate for $Q \bar Q$ at rest to decay to the final state. 
The relevant Feynman graphs are shown in Fig.~\ref{fig:upsdecay}. 
The decay rate $\Upsilon \to ggg$ is the inclusive decay rate for $\Upsilon
\to {\rm hadrons}$, and the decay rate for $\Upsilon \to \gamma g g$ is that
for $\Upsilon \to \gamma + {\rm hadrons}$. The probability to find $Q \bar Q$
at the origin, $\abs{\psi(0)}^2$, is sensitive to the detailed dynamics of the
$Q \bar Q$ bound state. If one takes the ratio of decay rates,
$\abs{\psi(0)}^2$ drops out, and the ratio of decay rates can be used to
determine $\alpha_s$.

The above qualitative discussion can be made precise using NRQCD
(non-relativistic QCD) to calculate the properties of the $\Upsilon$~\cite{BBL}. NRQCD has
an expansion in powers of $v$, the velocity of quarks in the bound state. The
$\alpha_s$ expansion is coupled to the $v$ expansion, since $v \sim \alpha_s$
in a Coulombic system. The NRQCD approach allows one to systematically factor
the decay rate into short distance coefficients that are calculable in
perturbation theory, and non-perturbative hadronic matrix elements that
generalize the notion of the wavefunction at the origin. The $\Upsilon$
wavefunction in NRQCD has different Fock components. The lowest order (in $v$)
component is $\ket{\bar Q Q}$ and the first correction contains a gluon,
$\ket{\bar Q Q g}$. The $\ket{\bar Q Q g}$ is referred to as the color-octet
component, because the two quarks are in a color octet state. The NRQCD velocity
counting rules show that the probability to find the $\Upsilon$ in the octet
component is of order $v^2$.

\def\darr#1{\raise1.5ex\hbox{$\leftrightarrow$}\mkern-16.5mu #1}

The decay rate for $\Upsilon \to \mu^+ \mu^-$ in NRQCD is~\cite{BBL}
\begin{equation}\label{51}
\Gamma(\Upsilon \to \mu^+ \mu^-) = {2 \im f_{ee}({}^3S_1)\over M^2_b}
\me{\Upsilon}{{\mathcal O}_1({}^3S_1)}{\Upsilon} +
{2 \im g_{ee}({}^3S_1)\over M^4_b}
\me{\Upsilon}{{\mathcal P}_1({}^3S_1)}{\Upsilon}
\end{equation}
where $M_b$ is the $b$-quark pole mass,
${{\mathcal O}_1({}^3S_1)}=\psi^\dagger\bsigma\chi
\cdot \chi^\dagger\bsigma\psi$ and ${{\mathcal P}_1({}^3S_1)}=
-i/4(\psi^\dagger\bsigma\chi
\cdot \chi^\dagger\bsigma (\darr{\bf D})^2 \psi + {\rm h.c.})$,
and $\psi$ and $\chi^\dagger$ annihilate quarks and antiquarks, respectively.
The first term is the leading order contribution, and the second term is the
$v^2$ correction. The coefficients $f_{ee}({}^3S_1)$ and $g_{ee}({}^3S_1)$
can be computed from the first graph in Fig.~\ref{fig:upsdecay} at lowest order
in perturbation theory. The values are~\cite{BBL}
\begin{equation}
\im f_{ee}({}^3S_1)={\pi \alpha^2 e_b^2\over 3}\left[1-16 {\alpha_s \over \pi}
\right],\qquad
\im g_{ee}({}^3S_1)=-{4\pi \alpha^2 e_b^2\over 9},
\end{equation}
where $e_b=-1/3$ is the charge of the $b$-quark, and the radiative correction to
$f_{ee}({}^3S_1)$ has also been included.

The decay rate for $\Upsilon \to {\rm hadrons}$ is~\cite{BBL}
\begin{eqnarray}
\Gamma(\Upsilon \to {\rm hadrons}) &=& {2 \im f_1({}^3S_1)\over M^2_b}
\me{\Upsilon}{{\mathcal O}_1({}^3S_1)}{\Upsilon} +
{2 \im g_1({}^3S_1)\over M^4_b}
\me{\Upsilon}{{\mathcal P}_1({}^3S_1)}{\Upsilon}\nonumber \\
&&+\Gamma^{(8)}(\Upsilon \to {\rm hadrons}) 
\end{eqnarray}
where $\Gamma^{(8)}$ is the contribution to the decay rate from the color-octet
component of the $\Upsilon$. In NRQCD, the color octet decay rate is $v^4$
suppressed relative to the color singlet decay rate. However, the color octet
component can decay into two gluons, rather than three, so the color octet
decay rate of order $\alpha_s^2 v^4$ can compete with the relativistic
correction to the color singlet decay rate of order $\alpha_s^3 v^2$.
The coefficient functions are~\cite{BBL}
\begin{eqnarray}
\im f_1({}^3S_1)&=& {10 \over 243}(\pi^2-9)\alpha_s^3(M_b)\nonumber\\
&&\times \left[1 + \left(
-9.46(2)C_F + 4.13 (17) C_A - 1.161(2) n_f){\alpha_s\over \pi}  \right)
\right]\nonumber \\
\im g_1({}^3S_1)&=& -{5 \over 1458}(19\pi^2-132)\alpha_s^3(M_b)
\end{eqnarray}
The $\alpha_s^3$ term for $\im f_1({}^3S_1)$ has the given value when the scale
of the $\alpha_s^2$ term is the $b$-quark pole mass.

The decay rate $\Gamma(\Upsilon \to \gamma+{\rm hadrons})$ has the form
\begin{equation}
\Gamma(\Upsilon \to \gamma+{\rm hadrons}) = {2 \im f_\gamma({}^3S_1)\over M_b^2}
\me{\Upsilon}{{\mathcal O}_1({}^3S_1)}{\Upsilon}+{2 \im g_\gamma({}^3S_1)\over M_b^2}
\me{\Upsilon}{{\mathcal P}_1({}^3S_1)}{\Upsilon},\label{53}
\end{equation}
where the coefficient function is~\cite{BBL}
\begin{eqnarray}
\im f_\gamma({}^3S_1)&=&{8\over 27}(\pi^2-9)\alpha_s^2(M_b)\alpha e_b^2
\nonumber \\ && \times
\left[1 + \left(
-9.46(2)C_F + 2.75 (11) C_A - 0.774(1) n_f){\alpha_s\over \pi}  \right)
\right]
\end{eqnarray}

The equations of motion can be used to relate the matrix elements of the
$S$-wave and $P$-wave operators~\cite{Gremm},
\begin{equation}
\me{\Upsilon}{{\mathcal P}_1({}^3S_1)}{\Upsilon}=
{M_\Upsilon-M_b\over M_b}
\me{\Upsilon}{{\mathcal O}_1({}^3S_1)}{\Upsilon}+O(v^2).\label{GK}
\end{equation}
The matrix element of the $S$-wave operator is the NRQCD analog of the
wavefunction at the origin in a potential model calculation. The matrix element
is non-perturbative, and can be eliminated by considering ratios of decay rates.
The matrix element of ${\mathcal P}_1({}^3S_1)$ can be determined from
Eq.~(\ref{GK}) using estimates of the $b$-quark pole mass $M_b$, and the
measured $\Upsilon$ mass, as was done in Ref.~\cite{Gremm}. However, one can
instead replace $(M_\Upsilon-2M_b)/M_b$ by $-4/9\alpha_s^2$, the lowest-order
result for the binding energy for a Coulomb bound state. This reduces somewhat
the uncertainty in the extraction of $\alpha_s$, since it eliminates any
uncertainty from the pole mass.

The experimental value of the ratio $\Gamma(\Upsilon \to {\rm
hadrons})/\Gamma(\Upsilon \to \ell^+ \ell^-)=39.11 \pm 0.4$~\cite{pdg}, where
$\ell=e,\mu,\tau$ gives $\alpha_s(M_b)=0.177\pm 0.01$, using the ratio of the
theoretical formula for the decay widths. The unknown octet decay rate has been estimated to be
less than 9\% in Ref.~\cite{Gremm}, and this has been included as a theoretical
uncertainty. The decay rates Eq.~(\ref{51})--(\ref{53}) have been written in
terms of $\alpha_s(M_b)$. One can instead rewrite them in terms of
$\alpha_s(\mu)$ using Eq.~(\ref{af}), extract $\alpha_s(\mu)$, and convert this
into $\alpha_s(M_b)$ using Eq.~(\ref{af}). We have include the uncertainty of a
scale change by a factor of two in the theoretical estimate. The octet and
scale uncertainties are comparable in size.

The experimental value of the ratio $\Gamma(\Upsilon \to \gamma + {\rm
hadrons})/\Gamma(\Upsilon \to {\rm hadrons})= 
2.75\pm0.04\pm0.15$~\cite{cleophot} can also be used to extract $\alpha_s$. It
is convenient to use the experimental value of $\Gamma(\Upsilon \to {\rm
hadrons})/\Gamma(\Upsilon \to \ell^+ \ell^-)$ to convert this to
$\Gamma(\Upsilon \to \gamma + {\rm hadrons})/\Gamma(\Upsilon \to \ell^+
\ell^-)=1.075\pm0.06$, before comparing with the theoretical results. This
eliminates the theoretical uncertainty due to the octet component in the
hadronic decay rate. The extracted value of $\alpha_s(M_b)$ is $0.189\pm0.01$,
where we have included a scale uncertainty as above.
Averaging the two extractions gives $\alpha_s(M_b)=0.183 \pm 0.01$
which corresponds to $\alpha_s(M_Z)=0.108 \pm 0.004$

\section{$\alpha_s$ FROM HADRON-HADRON SCATTERING}

There are many process at high-energy hadron-hadron colliders
which can constrain the value of $\alpha_s$. All rely on the QCD improved parton
model, and on the factorization theorems of QCD
\cite{factorization}. The rate for any process is expressed as a
convolution of the partonic scattering amplitude $\sigma_{i,j}^p$ and
parton distribution functions discussed in section \ref{sec:struc}; see 
Eq \ref{eq:parton} (note that here we use $f$ rather than $q$ as the
sum on $i,j$ runs over quarks and gluons. 
\begin{equation}
\sigma=\sum_{i,j}\int\! dx_1dx_2
f_i(x,M^2)f_j(x,M^2)\sigma_{ij}^p(M)
\end{equation}
The factorization scale $M$ is arbitrary. As in the case of the scale
$\mu$ used in $\alpha_s(\mu)$ (see Eq~\ref{eq:mu} and the surrounding
discussion), the exact result cannot depend on its choice. However as
the processes  $\sigma_{i,j}^p$ is only calculated to some finite order 
in perturbation theory, some residual $M$ dependence will remain. As
in the case of $\mu$ the sensitivity to $M$ will be small if it is
chosen to be a characteristic scale of the process; for example, in the 
case of the production of a pair of jets of momentum, $p_T$,
transverse to the direction defined by the incoming hadrons, $M=p_T$
is a reasonable choice.

The quantitative tests of QCD and the consequent extraction of
$\alpha_s$ which appears in $\sigma_{i,j}^p$ are possible only if the
process in question has been calculated beyond leading order in QCD
perturbation theory. 
The production of hadrons with large transverse momentum in
hadron-hadron collisions provides a direct probe of the
scattering of quarks and gluons:  $qq \to qq$,
$qg \to qg$, $gg \to gg$, {\it etc.}. Here the leading order term in
  $\sigma_{i,j}^p$ is of order $\alpha_s^2$ so the rates are sensitive
  to its value.
Higher--order QCD calculations of the jet
rates\cite{Ellis91}\ and
shapes are in impressive agreement with data\cite{Abe96}. This agreement
has led to the proposal that these data could be used to provide a 
determination of 
$\alpha_s$\cite{giele96}. A  set of structure functions is assumed and
Tevatron collider 
 data are fitted over a very large range
 of transverse momenta, to the QCD prediction for the underlying
 scattering process that depends on $\alpha_s$. The evolution of the 
coupling over this energy range (40 to 250 GeV) is therefore tested in the
 analysis.
CDF obtains 
 $\alpha_s(M_Z)=0.1129\pm 0.
 0001~(\hbox{stat.}) \pm 0.
 0085~(\hbox{syst.}) $ \cite{cdfalpha}. Estimation of the theoretical errors
 is not straightforward. The structure functions used depend implicitly on
 $\alpha_s$ and an iteration procedure must be used to obtain a consistent
 result; different sets of structure functions yield different correlations
 between the two values of $\alpha_s$.
 We estimate an uncertainty of $\pm 0.005$ from examining the fits.
 Ref. \cite{giele96} estimates the error from unknown higher order QCD 
 corrections to be $\pm 0.005$. Combining these then gives:
$\alpha_s(M_Z)=0.1129\pm 0.011 $

QCD corrections to Drell-Yan type
cross sections (the production in hadron collisions
by quark-antiquark annihilation
of lepton pairs of invariant mass $Q$
from virtual photons or of  real $W$ or $Z$~bosons), are known\cite{Altarelli78}.
These processes are not very sensitive to $\alpha_s$ as the leading
piece in $\sigma_{i,j}^p$ is of order $\alpha_s^0$.
The production of $W$ and $Z$~bosons and photons at large
transverse momentum  begins at  order $\alpha_s^0$.
 The leading-order QCD subprocesses are $q\overline{q}  \to \gamma g$
and $qg \to \gamma q$.
The next-to-leading-order  QCD corrections are
known\cite{Aurence90}\cite{Baer90}\ 
for photons, and  for $W/Z$ production\cite{Baer91},
 and so an extraction of $\alpha_s$ 
is possible in principle.

 Data exist on photon production from the
CDF and D\O\ collaborations\cite{cdfgamma}\cite{d0gamma} and from 
fixed target experiments \cite{e706}. Detailed comparisons 
with QCD predictions
\cite{owens99} may indicate an excess of the data over the theoretical 
prediction at low value of transverse momenta, although other authors
\cite{vogelsang} find smaller excesses. These differences indicate
that while the process may be understood, no meaningful extraction of
$\alpha_s$ is possible.

The UA2 collaboration\cite{ua292} has extracted a value 
of  $\alpha_s(M_W)=0.123\pm 0.018(\hbox{stat.})\pm 0.017(\hbox{syst.})$
from the measured ratio
 $R_W={\sigma(W+1\hbox{jet})}/{\sigma(W+0\hbox{jet})}$.
The result depends on the algorithm used to define a jet,
and the dominant systematic errors due to fragmentation and corrections for 
underlying events 
(the former causes jet energy to be lost, the latter causes
it to be increased) are
connected to the algorithm.
The scale at which $\alpha_s(M)$ is to be evaluated
is not clear. A change from $\mu=M_W$ to $\mu=M_W/2$ causes a
 shift of 0.01 in the
extracted $\alpha_s$, and
the quoted error should be increased to take this into
account. 
There is also dependence on the 
parton distribution functions, and hence, $\alpha_s$ appears
explicitly in the formula for 
$R_W$, and implicitly in the distribution functions.
 Data from CDF and D\O\ on the W $p_T$
 distribution \cite{d0w} are in
 agreement with QCD but are not able to determine $\alpha_s$ with 
sufficient precision to have any weight in a global average.

The production rates of $b$ quarks in $p\overline{p}$ have been used to determine $\alpha_s$
\cite{geiser}. The next to leading order QCD production processes
 \cite{mangano} have been used. At order $\alpha_s$ the production
 processes are $gg\to b\overline{b}$ and
 $q\overline{q}\to b\overline{b}$ result in b-hadrons that are back to
 back in azimuth. By selecting events in this region the next-to
 leading order calculation can be used to compare rates to the
 measured value and a value of  $\alpha_s$ extracted. The errors are
 dominated by  
 the measurement errors, the choice of $\mu$ and $M$, and uncertainties 
 in the choice of structure functions. The last were estimated by
varying the structure functions used. The result is
$\alpha_s(M_Z)=0.113^{+0.009}_{-0.013}$.

\section{CONCLUSION}

\begin{figure}
\dofig{4in}{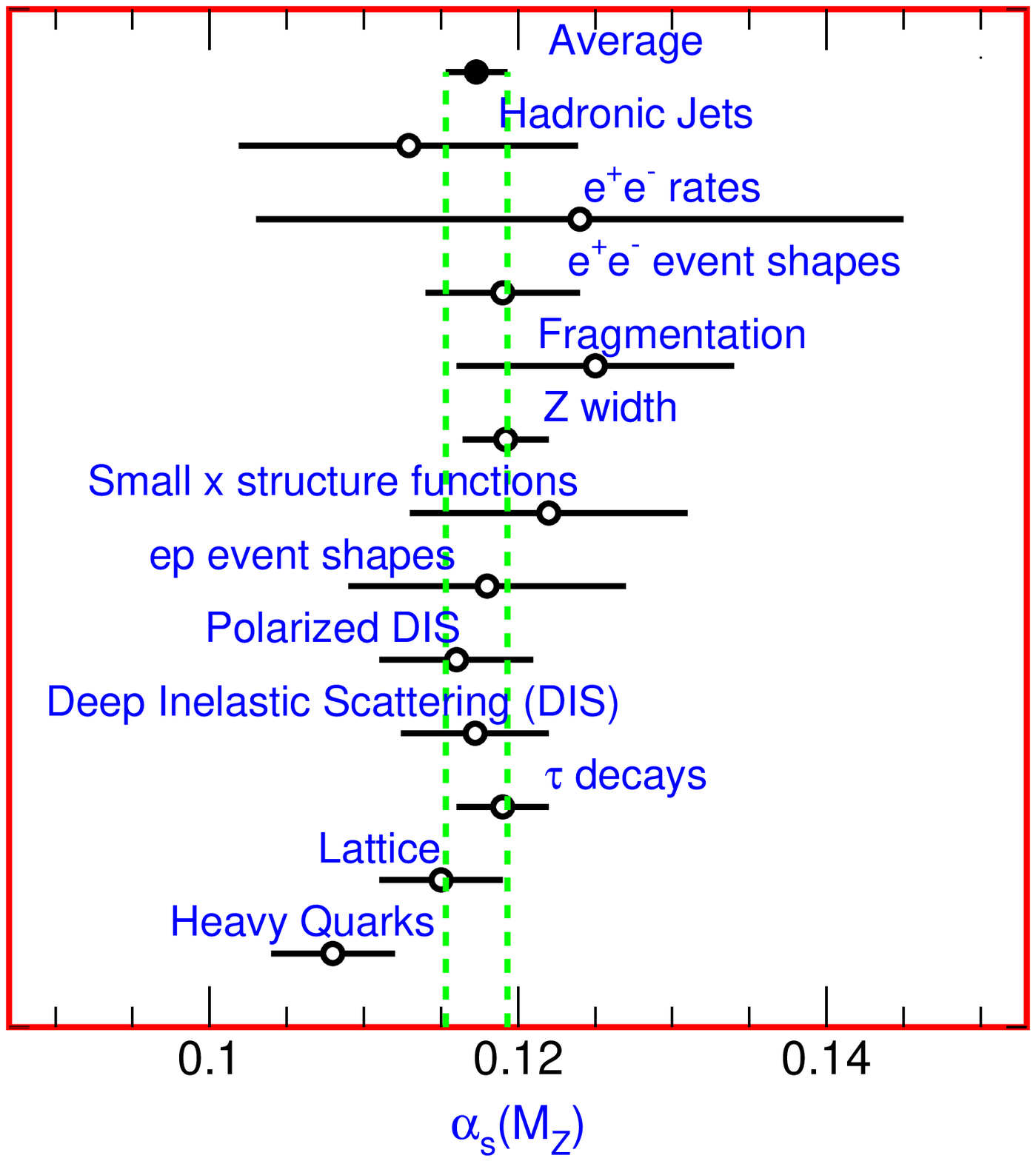}
\caption{Summary of $\alpha_s(M_Z)$}
\label{figalphas}
\end{figure}

\begin{figure}
\dofig{4in}{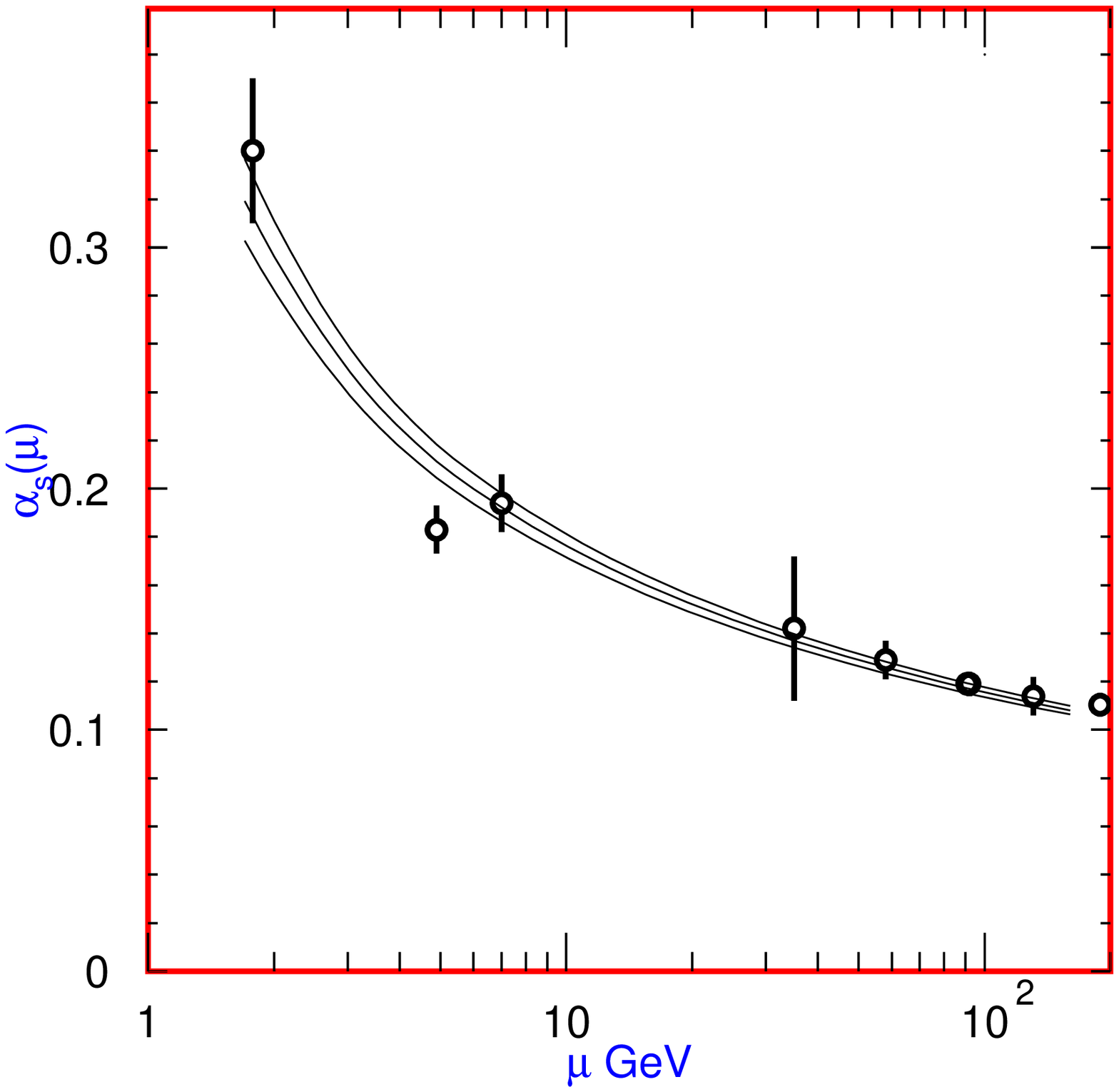}
\caption{Values of  $\alpha_s(\mu)$ and the values of $\mu$ where they 
  are measured. these results are, in increasing value of $\mu$,
 $\tau$ width, heavy quark decays, deep inelastic at $\sqrt{<Q^2>}=7$
 GeV, 
$e^+e^-$ annihilation rate at 35 
 GeV,  $e^+e^-$ event shapes at 58 GeV, the hadronic $Z$
width, $e^+e^-$  event shapes at Z,   130 and
at 189 GeV. The lines show our average  and $1\sigma$ errors. }
\label{runfigalpha}
\end{figure}

The previous sections have illustrated the large number of processes
where quantitative tests of QCD can be made and a value of $\alpha_s$
extracted.  
Figure~\ref{figalphas}
shows the values of $\alpha_s(M_Z)$ deduced from the
various processes shown above. The consistency and precision of these 
results is remarkable. Figure~\ref{runfigalpha}
 shows the values of $\alpha_s(\mu)$ and the
values of $\mu$ where they are measured. This figure clearly shows the
experimental evidence for the variation of $\alpha_s(\mu)$ with $\mu$
predicted by Eq.\ref{3}. 

An average
of the values in Figure~\ref{figalphas} and in Table~\ref{tab1}
 gives $\alpha_s(M_z)=0.1173$, with a total
$\chi^2$ of 9 for twelve  fitted points, showing good consistency among 
the data. The value from heavy quark systems contributes slightly more 
that one half of the total $\chi^2$. If this result is omitted the
average increases to 0.1185. All of the other results are within
$1.1\sigma$ of the average value.
The error on the average,
assuming that all of the errors in the contributing 
results are uncorrelated, is $\pm 0.0014$, and
may be an underestimate. We have seen that in almost 
all of  the cases discussed, the errors are dominated by systematic,
 usually theoretical errors. Only some of these, 
notably from the choice of scale, are correlated.
it is important to note that the average is not dominated by a 
single measurement;
there are many results with comparable small errors: 
from $\tau$~decay, lattice gauge, theory deep 
inelastic scattering and the $Z^0$ width.
We quote our average value  as
$\alpha_s(M_Z)=0.1173\pm 0.002$, which 
corresponds to $\Lambda^{(5)}=200 ^{+24} _{-23}$ MeV
using Eq.~\ref{6}.  The reader may wish to consult other recent
articles for different opinions\cite{alpha98,womersley}.

Significant improvements in the precision in the near future are not
likely.
 The
accuracy of data from LEP will not improve. It is possible  that a better
understanding of the jet rates in hadron-hadron colliders and a systematic
treatment of the errors from the structure  functions will lead to and
improvement in the precision of the value  of $\alpha_s$ derived. In many cases
where the data are quite precise,  such as heavy quark system, theoretical
uncertainties  limit the precision. In the very long term precision at the 1\%
level may be achievable\cite{burrows97}.

\section*{Acknowledgments}
 This work was supported by the Director, Office of Science, Office
of Basic Energy Services, of the U.S. Department of Energy under
Contract DE-AC03-76SF0098, and by DOE grant DOE-FG03-97ER40546.
\begin{table}
\begin{tabular}{|c|c|}
\hline
Process & $\alpha_s(M_z)$\\
\hline
$Z$ width & $0.1192 \pm 0.0028$\\
$e^+e^-$ rate at $\sqrt{s}=34$ GeV  & $0.124 \pm 0.021$\\
Structure functions  & $0.1172 \pm 0.0045$\\
Small $x$ structure functions  & $0.122 \pm 0.009$\\
Spin  structure functions  & $0.116 \pm 0.006$\\
Fragmentation functions & $0.125 \pm 0.009$\\
Event Shapes in $e^+e^-\to X$ & $0.119 \pm 0.005$\\
Event Shapes  in $ep\to eX$ & $0.118 \pm 0.009$\\
Tau decay & $0.119 \pm 0.003$\\
Lattice &$0.115 \pm 0.004$\\
Heavy quark systems & $0.108 \pm 0.004$\\
Jets in $pp$ & $0.129 \pm 0.011$\\
\hline
\end{tabular}
\caption{Values of $\alpha_s(M_z)$ used in the average}
\label{tab1}
\end{table}


\begin{thebibliography}{299} 
%
\bibitem{politzer}H.~D.~Politzer,
\prl30,1346 (1973); D.~J.~Gross and F.~Wilczek,
\prD9,980(1974)
\bibitem{theta}
R.~D.~Peccei and H.~R.~Quinn,
\prl38,1440(1977); S. Weinberg, \prl37,657(1977)
\bibitem{msbar}W.A.~Bardeen \etal, \prD18,3998(1978)
\bibitem{choice}G.~Grunberg, \pl95B,70(1980);
 \prD29,2315(1984);P.M.~Stevenson, \prD23,2916(1981);
 and 
\npB203,472(1982);
 S. Brodsky and  H.J. Lu,
SLAC-PUB-6389 (Nov.\ 1993); S. Brodsky, G.P. Lepage, and P.B. Mackenzie, 
\prD28,228(1983)
\bibitem{larin97}S.A. Larin, T. van Ritbergen, and
J.A.M. Vermaseren, \plB400,379(1997)
\bibitem{chetyk97}%
K.~G.~Chetyrkin, B.~A.~Kniehl and M.~Steinhauser,
\npB510,61(1998)
\bibitem{ope} K. Wilson \pr179,1499 (1969)
\bibitem{lautrup} B.~Lautrup, \plB69,109(1977)
\bibitem{thooft} G. 't Hooft, in {\sl The Whys of Subnuclear Physics}, ed. by A.Zichichi, (Plenum, New York, 1978)
\bibitem{david} F. David, \npB209,433(1982), \npB234,237(1984)
\bibitem{bsuv} I.I.~Bigi, M.~Shifman, N.G.~Uraltsev and A.~Vainshtein,
\prD50,2234(1994)
\bibitem{beneke} M.~Beneke and V.~Braun, \npB426,301(1994)
\bibitem{ren-review} M.~Beneke,\prept317,1(1999)
\bibitem{KLN} T.~D.~Lee and M.~Nauenberg,
\pr133,B1549 (1964); T.~Kinoshita,
J.\ Math.\ Phys.\  {\bf 3} 650 (1962)
\bibitem{akhoury}
A.~Sinkovics, R.~Akhoury and V.~I.~Zakharov, \prD58,114025(1998)
\bibitem{Gorishny91}S.G.~Gorishny, A.~Kataev,
and S.A.~Larin, \plB259,114(1991)
L.R. Surguladze and M.A. Samuel,
\prl66,560(1991)
\bibitem{kuhn94} K.G. Chetyrkin and
J.H. Kuhn, 
\plB308,127(1993) 
\bibitem{cleoR}R. Ammar \etal, 
\prD57,1350(1998) 
\bibitem{haidt}D. Haidt, in 
{\it Directions in High Energy Physics,} vol.~14, p.~201,
ed.\ P. Langacker (World Scientific, 1995)
\bibitem{lepz}G. Quast,
presented at the European Physical Society Meeting, Tampere, Finland (July 1999)
%
\bibitem{blondel93}A. Blondel and C. Verzegrassi,
\plB311,346(1993)
G. Altarelli \etal, \npB405,3(1993)
\bibitem{langsports} J. Erler and  P Langacker, 
Brief review in Eur.Phys.J.C3:1-794,1998. 
\bibitem{Altarelli77}V.N. Gribov and 
L.N. Lipatov, \sjnp15,438(1972)
Yu.L. Dokshitzer, 
\jetp46,641(1977)
\bibitem{altpar77} 
G.~Altarelli and G.~Parisi, \npB126,298(1977)
\bibitem{Curci80}G.~Curci, W.~Furmanski, and 
R.~Petronzio, \npB175,27(1980)
%
R.~Petronzio, \pl97B,437(1980);
 and \zpC11,293(1982)
E.G.~Floratos, 
C.~Kounnas, and R.~Lacaze, \pl98B,89(1981);
\pl98B,285(1981); and \npB192,417(1981)
R.T.~Herrod and 
S.~Wada, \pl96B,195(1981); and \zpC9,351(1981)
\bibitem{Gluck82}M.~Gl\"uck \etal, \zpC13,119(1982)
\bibitem{sterman}G. Sterman, \npB281,310(1987)
S. Catani and L. Trentadue, \npB327,323(1989);
\np B353,183(1991)
\bibitem{gross}D. Gross and  C.H. Llewellyn Smith, \npB14,337(1969)
\bibitem{kataev}J. Chyla and A.L. Kataev, \plB297,385(1992)
\bibitem{larinj}
S.A. Larin and J.A.M. Vermaseren, \plB259,345(1991)
%
\bibitem{kataev1} A.L. Kataev and V.V. Starchenko,  \mplA10,235(1995)
\bibitem{braun}V.M. Braun and A.V. Kolesnichenko, \npB283,723(1987)
\bibitem{dasgupta}M. Dasgupta and B. Webber, \plB382,273(1993)
\bibitem{ccfrr98}J. Kim \etal, \prl81,3595(1998)
\bibitem{f3rest}D. Allasia \etal, \zpC28,321(1985)
K. Varvell \etal, \zpC36,1(1997)
V.V. Ammosov \etal, \zpC30,175(1986) 
P.C. Bosetti \etal, \npB142,1(1978)
\bibitem{parente}A.L. Kataev \etal, \tt hep-ph9907310\rm
\bibitem{Whitlow}L.W. Whitlow \etal, \plB282,475(1992)
\bibitem{Benvenuti}A.C.~Benvenuti \etal, 
\plB223,490(1989);
\plB223,485,(1989);
\plB237,592(1990); and
\plB237,599(1990)
\bibitem{adams}M.R. Adams \etal,  \prD54,3006(1996)
\bibitem{hera-dis}M. Derrick \etal, \plB345,576(1995); \zpC62,399(1999)
\bibitem{santiago}J. Santiago and F.J. Yndurain, \tt hep-ph/9904344\rm
\bibitem{moments}K. Adel, F. Barriero, and F.J. Yndurain, 
\npB495,221(1997)
W.L. Van Neerven, and E.B. Zijlstra, \plB272,127(1991);
\npB383,525(1992)
S. A. Larin \etal, \npB427,41(1994);
\npB492,338(1997)
\bibitem{Bazizi91}K. Bazizi and S.J. Wimpenny, UCR/DIS/91-02 
\bibitem{Virchaux92}M. Virchaux and  A. Milsztajn, 
\plB274,221(1992)
\bibitem{Quintas93}P.Z. Quintas, \prl71,1307(1993)
\bibitem{abeabe}K. Abe \etal,  \prl74,346(1995);
\plB364,61(1995);
\prl75,25(1995)
\bibitem{smc98}B. Adeva \etal,  \prD58,112002(1998)
\bibitem{spinsf}D. Adams \etal, \plB329,399(1995);
\prD56,5330(1998);
\prD58,1112001(1998)
\bibitem{hermes}A. Airapetian \etal,  \plB442,484(1998),
 \tt hep-ex/99-06035\rm
\bibitem{bjsum}J.D. Bjorken, \pr148,1467(1966)\bibitem{Mertig:1996ny}
R.~Mertig and W.~L.~van Neerven,
\zpC70,637(1996)
%
\bibitem{altbj}G. Altarelli \etal, \npB496,337(1997), \tt
  hep-ph/9803237\rm
\bibitem{ellisbj} J. Ellis \etal, \prD54,6986(1996)
\bibitem{bfkl}
A. DeRujula \etal, \prD10,1669(1974)
E.A. Kurayev, L.N. Lipatov, and V.S. Fadin, 
\jetp45,119(1977).
Ya.Ya. Balitsky and L.N. Lipatov,
\sjnp28,882(1978)
\bibitem{ball}R.D. Ball and S. Forte, \plB335,77(1994);
\plB336,77(1994)
H1\rm Collaboration: S. Aid \etal, \npB470,3(1996)
\bibitem{Ball96}
R.D. Ball and A. DeRoeck, \tt hep-ph/9609309\rm
European Physical Society meeting, Brussels, (July 1995)
\bibitem{h194}\bf H1\rm\ Collaboration: T. Ahmed \etal,
  \npB439,471(1995)
\bibitem{catha}S. Catani and F. Hautmann, \npB427,475(1994)
\bibitem{partonfits}H.L. Lai \etal,  \tt hep-ph/9903282\rm
 A.D. Martin \etal, \tt hep-ph/9906231\rm  
\bibitem{witten77} E. Witten, \npB120,189(1977)
\bibitem{but99} J. Butterworth,  {\it International Conference on 
Lepton Photon Interactions}, Stanford, USA  (Aug.\ 1999)
%
\bibitem{ackerstaff97}K. Ackerstaff \etal, \plB412,225(1997);
\plB411,387(1997) 
P. Abreu \etal, \zpC69,223(1996)
R. Barate \etal, \plB458,152(1999)
M. Acciarri \etal, \plB436,403(1998), L3 preprint
  204 (2000)
\bibitem{muramatsu94}K. Muramatsu \etal, \plB332,477(1994)
\bibitem{sahu95} S.K. Sahu \etal, \plB346,208(1995)
\bibitem{h1gamma}\bf  H1\rm\ Collaboration: C. Adloff \etal, DESY-98-205
J.Breitweg \etal,  DESY 99-057
\bibitem{frixone96} S. Frixone,  \npB507,295(1997)
B.W. Harris and J.F. Owens,  \prD56,4007(1997)
M. Klasen and G. Kramr, \zpC72,107(1996)
\bibitem{webber95}P. Nason and B.R. Webber, \npB421,473(1994)
\bibitem{alephfrag}D. Buskulic \etal, 
\plB357,487(1995)
{\it ibid.,} erratum \plB364,247(1995)
\bibitem{opalfrag}\bf OPAL\rm\ Collaboration: R. Akers \etal, \zpC68,203(1995)
\bibitem{delphifrag}\bf DELPHI\rm\ Collaboration: P. Abreu \etal,
  \plB398,194(1997)
\bibitem{Farhi77}E.~Farhi, \prl39,1587(1977)
\bibitem{Basham78}C.L.~Basham \etal,
\prD17,2298(1978)
\bibitem{Andersson83}B.~Andersson \etal, \prept97,33(1983)
%
A.~Ali \etal, \npB168,409(1980)
A.~Ali and R.~Barreiro, \pl118B,155(1982)
%
B.R.~Webber, \npB238,492(1984)
G. Marchesini \etal, Phys.\ Comm.\ {\bf 67}, 465 (1992)
%
T. Sjostrand and M. Bengtsson,
\cpc43,367(1987)
T. Sjostrand, CERN-TH-7112/93 (1993)
\bibitem{Bethke88}S. Bethke \etal, \plB213,235(1988)
%
\bibitem{durham}S. Bethke \etal, \npB370,310(1992)
%
\bibitem{Akrawy91}M.Z. Akrawy \etal, \zpC49,375(1991)%
\bibitem{fixedAbe95}K. Abe \etal, \prl71,2578(1993); \prD51,962(1995)
\bibitem{opalshape}P.D. Acton \etal, \zpC55,1(1992);
\zpC58,386(1993)
\bibitem{l3shape}O. Adriani \etal, \plB284,471(1992)
\bibitem{alephshape}D. Decamp \etal, \plB255,623(1992);
 \plB257,479(1992)
\bibitem{e-jet}J. Ellis, M.K. Gaillard, and G. Ross, \npB111,253(1976)
    {\it ibid.}, erratum  \npB130,516(1977)
 P. Hoyer \etal,
    \npB161,349(1979)
\bibitem{Ellis80}R.K. Ellis, D.A. Ross, T. Terrano, \prl45,1226(1980)
Z. Kunszt and P. Nason, ETH-89-0836 (1989)
\bibitem{opal}O. Adriani \etal,
\plB284,471(1992)
M. Akrawy \etal, \zpC47,505(1990)
B. Adeva \etal, \plB248,473(1990) 
%
D. Decamp \etal, \plB255,623(1991)
\bibitem{dok95}Y.L. Dokshitzer and B.R. Webber \plB352,451 (1995)
Y.L. Dokshitzer {\it et. al.}
\npB511,396 (1997)
Y.L. Dokshitzer {\it et. al.} JHEP 9801,011 (1998)
\bibitem{delphi99}
\bf DELPHI\rm\   Collaboration: \plB456,322(1999)
\bibitem{catani} S. Catani \etal, \plB263,491(1991)
%
\bibitem{webber93} S. Catani \etal, \plB269,432(1991)
S. Catani, B.R. Webber, and G. Turnock, \plB272,368(1991)
N. Brown and J. Stirling,
\zpC53,629(1992)
\bibitem{matching}
G. Catani \etal, \plB269,632(1991);
\plB295,269(1992);
\npB607,3(1993);
\plB269,432(1991)
\bibitem{topaz93}Y. Ohnishi \etal, \plB313,475(1993)
\bibitem{cleoshape} L. Gibbons \etal, CLNS 95-1323 (1995)
\bibitem{lep130}\bf DELPHI\rm\ Collaboration: D. Buskulic \etal,
\zpC73,409(1997);
\zpC73,229(1997)
\bibitem{lep190}\bf  ALEPH\rm\ Collaboration: 99-023 (1999);
\bf DELPHI\rm\ Collaboration: 99-114 (1999); \bf L3\rm\ Collaboration:
 L3-2414 (1999);
\bf OPAL\rm\ Collaboration, PN-403 (1999); all submitted to {\it International
Conference on
Lepton Photon Interactions},\rm Stanford, USA (Aug.\ 1999)
\bibitem{*acc}
\bf OPAL\rm\ Collaboration: M. Acciarri \etal, \plB371,137(1996);
\zpC72,191(1996)
K. Ackerstaff \etal, \zpC75,193(1997)
\bf ALEPH\rm\   Collaboration: ALEPH 98-025 (1998)
\bibitem{opaljade}\bf JADE \rm\  and \bf OPAL \rm\ Collaborations
  CERN/EP-99-175.
\bibitem{delphishape}P. Abreu \etal, \zpC59,21(1993),  CERN-EP/99-133
\bibitem{graudenz}D. Graudenz, \prD49,3921(1994)
\tt hep-ph/9708362\rm
J.G. Korner, E. Mirkes, and G.A. Schuler, 
\ijmpA4,1781,(1989)
S. Catani and M. Seymour, \npB485,291(1997)
M. Dasgupta and B.R. Webber,
\tt hep-ph/9704297\rm
E. Mirkes and D. Zeppenfeld, \plB380,205(1996)
%
\bibitem{h195}\bf  H1\rm\ Collaboration: T. Ahmed \etal, \plB346,415(1995);
\epjC5,575(1998)
%
\bibitem{zeus95}\bf ZEUS\rm\ Collaboration: M. Derrick \etal,
\plB363,201(1995) 
\bibitem{pich}S. Narison and A. Pich, \plB211,183(1988)
\bibitem{Braatennar}
 E. Braaten, S. Narison,
and A. Pich, \npB373,581(1992)
\bibitem{braaten}
E. Braaten, \prl60,1606(1988)
\bibitem{cleotau}T. Coan \etal (CLEO Collaboration), \plB356,580(1995)
\bibitem{alephtau} ALEPH Collaboration, \epjC4,409(1998)
\bibitem{opaltau} OPAL Collaboration, \epjC7, 571 (1999)
\bibitem{gaugnelli}G. de Divitiis \etal, \npB437,447(1995)
\bibitem{mlus} M. Luscher \etal, \npB413,481(1994)
\bibitem{Khadra96} J. Shigemitsu, 
Nucl.\ Phys.\ {\bf B} (Proc.\ Supp.) {\bf 53}, 16 (1997)
\bibitem{davies97}C.T.H.  Davies \etal, \prD56,2755(1997)
\bibitem{sesam99}
A.~Spitz {\it et al.}  [SESAM Collaboration],
\prD60,074502(1999)
\bibitem{BBL} G.~T.~Bodwin, E.~Braaten and G.~P.~Lepage,
\prD51,1125(1995)
\bibitem{Gremm}
M.~Gremm and A.~Kapustin, \plB407,323(1997)
\bibitem{pdg} Particle Data Group, \epjC3,1(1998)
\bibitem{cleophot} B. Nemati \etal, \prD55,5273(1997)
\bibitem{factorization}  A. H. Mueller, \prD18,3705(1978)
 J. Collins, D. Soper
and G. Sterman, \npB223,81(1983), \pl109B,388(1983)
 R. K. Ellis,
 \npB152,285(1979).
Collins JC, Soper DE, Sterman G. in 
{\it Perturbative Quantum Chromodynamics}, edited
 by A. H. Mueller (World Scientific Singapore), p. 1.
\bibitem{Ellis91}S.D. Ellis, Z.~Kunszt,
and D.E.~Soper, \prl64,2121(1990)
F. Aversa \etal, \prl65,401(1990)
W.T. Giele, E.W.N. Glover, and D. Kosower, \prl73,2019(1994)
S. Frixione, Z. Kunszt, and  A. Signer,
\npB467,399(1996) 
\bibitem{Abe96}F. Abe \etal,  \prl77,438(1996)
B. Abbott \etal, \prl82,2451(1999)
\bibitem{giele96}  W.T. Giele, E.W.N. Glover, and J. Yu,
\prD53,120(1996) 
\bibitem{cdfalpha}\bf   CDF\rm\ Collaboration reported in \cite{womersley}
%
\bibitem{Altarelli78}G.~Altarelli, R.K.~Ellis, and G.~Martinelli,
\npB143,521(1978)
\bibitem{Aurence90}P. Aurenche, R. Baier, and M. Fontannaz,
\prD42,1440(1990)
P.~Aurenche \etal, \pl140B,87(1984)
P.~Aurenche \etal, \npB297,661(1988)
\bibitem{Baer90}H. Baer, J.~Ohnemus, and J.F.~Owens, \plB234,127(1990)
\bibitem{Baer91}H. Baer and  M.H.~Reno, \prD43,2892(1991);
P.B. Arnold and M.H. Reno, \npB319,37(1989)
\bibitem{cdfgamma}F. Abe \etal, \prl73,2662(1994)
\bibitem{d0gamma}S. Abachi \etal, 
       \prl77,5011(1996)
\bibitem{e706}G. Alverson \etal,
\prD48,5(1993) 
\bibitem{owens99}L. Apanasevich \etal,
\prD59,074007(1999);
\prl81,2642(1998) 
\bibitem{vogelsang}W. Vogelsang and  A. Vogt,
\npB453,334(1995) 
P. Aurenche \etal, 
\epjC9,107(1999) 
\bibitem{ua292}J. Alitti \etal, \plB263,563(1991)
\bibitem{d0w} 
S. Abache \etal,
\prl75,3226(1995);
J. Womersley, private communication
J. Huston, in the \it Proceedings to the
29th International Conference on High-Energy Physics (ICHEP98)\rm,
Vancouver, Canada (23--29 Jul 1998)
\tt hep-ph/9901352 \rm
\bibitem{d0resum} D0 Collaboration: B. Abbott \etal,
 \tt hep-ex/9907009\rm
T. Affolder \etal,  FERMILAB-PUB-99/220
\bibitem{mangano}
M.~L.~Mangano, P.~Nason and G.~Ridolfi,
\npB373,295(1992)
\bibitem{geiser} C. Albajar et al. \plB369,46(1996) 
\bibitem{alpha98}For example see, 
S. Bethke, in \it Proceedings of the IV$^{th}$ Int.\ Symposium on 
Radiative Corrections\rm, 
Barcelona, Spain (Sept.\ 1998),
\tt hep-ex/9812026\rm
M Davier, \it 33rd Rencontres de Moriond: Electroweak Interactions and Unified Theories\rm , Les Arcs, France (14--21 Mar.\ 1998)
P.N. Burrows, Acta.\ Phys.\ Pol.\ {\bf28}, 701 (1997)
\bibitem{womersley}J. Womersley, {\it International Conference on 
Lepton Photon Interactions}, Stanford, USA  (Aug.\ 1999) 
%
\bibitem{burrows97}P.N. Burrows \etal, in {\it
Proceedings of 1996 DPF/DPB Snowmass Summer Study\rm},
ed.\ D. Cassel \etal, (1997)

\end{thebibliography}
\end{document}